\documentclass[reprint,aps,prc,twocolumn,superscriptaddress,floatfix,amsmath,amssymb,10pt]{revtex4-2}
\DeclareMathAlphabet{\mathpzc}{OT1}{pzc}{m}{it}
\usepackage{bm}
\usepackage{dcolumn}
\usepackage{amsthm}
\usepackage{amsmath}
\usepackage{amssymb}
\usepackage{graphicx}
\usepackage{xcolor}
\usepackage{fix-cm}
\usepackage{mathptmx} 
\usepackage[T1]{fontenc}
\usepackage[colorlinks,allcolors=blue]{hyperref}
\makeatletter
\def\NAT@def@citea{\def\@citea{\NAT@separator}}
\makeatother

\begin{document}

\title{Symmetry breaking and restoration on a fermionic quantum ring}

\author{Joshua Cesca}
\affiliation{Department of Fundamental and Theoretical Physics, Research School of Physics, The Australian National University, Canberra ACT 2601, Australia}
\author{C\'edric Simenel}\email{cedric.simenel@anu.edu.au}
\affiliation{Department of Fundamental and Theoretical Physics, Research School of Physics, The Australian National University, Canberra ACT 2601, Australia}
\affiliation{Department of Nuclear Physics and Accelerator Applications, Research School of Physics, The Australian National University, Canberra ACT 2601, Australia}

\date{\today}

\begin{abstract}
\edef\oldrightskip{\the\rightskip}
\begin{description}
\rightskip\oldrightskip\relax
\setlength{\parskip}{0pt} 
\item[Background] The Hartree-Fock mean-field approximation is standard in combination with energy density functionals (EDF) that account for some dynamical correlations. Breaking and restoring the symmetries of the system allow for the inclusion of additional static correlations. However, exact solutions  to evaluate the effectiveness of these methods are rare.
\item[Purpose] To benchmark the Hartree-Fock method with broken and restored rotational symmetry in a system of identical interacting fermions on a one-dimensional quantum ring using model interactions.
\item[Method] The ground-state wave function is found using the Hartree-Fock method both with rotational invariance and with the symmetry broken at the mean-field level. Rotational symmetry is then restored
with an angular momentum projection method. The ground-state energies are compared to variational Monte Carlo predictions. This is done for a range of different interactions between the particles.
\item[Results] Breaking the rotational symmetry in the Hartree-Fock mean-field brings little improvement to the ground-state energy in weakly repulsive systems or attractive systems confined on small rings. Larger improvements are found in strongly repulsive systems and attractive systems on larger rings in which the particles form a self-bound system. Symmetry restoration brought only small improvements in most cases but was able to account for most of the remaining correlation energy (after symmetry breaking) in repulsive systems.
\item[Conclusions] The effectiveness of incorporating correlations through rotational symmetry breaking followed by angular momentum projection is demonstrated for one-dimensional quantum rings using model interactions, encouraging generalisations to other symmetries, extensions to higher dimensions, as well as applications in the EDF framework. 
\end{description}
\end{abstract}

\maketitle


\section{Introduction}

Realistic descriptions of quantum many-fermion systems are often challenging and the Hartree-Fock (HF) mean-field approach is often used as a first approximation. 
Although the HF equation can be obtained directly from the interaction between particles, in the nuclear physics context they are usually derived from an energy density functional (EDF), thus accounting for some crucial correlations. 
Such EDF approaches have had a large degree of success in nuclear structure \cite{benderSelfconsistentMeanfieldModels2003,peru2014} and reaction \cite{simenel2012,simenel2018} studies.

The interaction between nucleons is invariant under a number of symmetries, such as translational, rotational and global U(1) gauge invariances. 
Exact solutions to the many-body Schr\"{o}dinger equation are also expected to have the same symmetries as the interactions. 
However, mean-field approaches often work in bases which do not respect the symmetries of the system and thus allow for symmetry-broken solutions. 
This feature of mean-field methods 
provides a way for the method to account for some additional correlations between particles while still working in the simple independent-particle picture.

The spontaneous symmetry breaking of the mean-field solutions does come at a price. From Noether's theorem, each continuous symmetry of a system is associated with a conservation law for that system. As a result, the mean-field symmetry-broken solutions are no longer states with well-defined values of the relevant conserved quantity, affecting 
calculations of physically relevant quantities. For instance, the electromagnetic transition probabilities between nuclear states are subject to selection rules which depend on the parity and total angular momentum of the initial and final states. Not having good quantum numbers in the symmetry-broken states makes these calculations difficult.

The symmetry restoration method \cite{sheikhSymmetryRestorationMeanfield2021} offers a solution to this problem. 
The method projects the symmetry-broken state back onto a state with a 
good quantum number. It does this by taking an appropriately weighted superposition of each of the states in the broken symmetry group. This superposition results in a state with the correct symmetry while still keeping the symmetry-broken solution's account of correlations.
Symmetry restoration methods are then commonly used in various fields. 
See, e.g.,  applications within the coupled cluster theory in quantum chemistry \cite{qiu2017a,qiu2017b} and in nuclear physics \cite{duguet2015,duguet2017,hagen2022}, as well as HF applications with shell model interactions \cite{lauber2021} and with Skyrme and Gogny functionals \cite{benderSelfconsistentMeanfieldModels2003,robledo2019}. 

Model systems are often used to benchmark approximate methods as they are simple enough to be solved quasi-exactly, yet still share important features with real physical systems. 
Infinite nuclear matter (see, e.g., \cite{leonhardt2020}) is one system which has been used to study the binding energy of nuclei \cite{atkinsonReexaminingRelationBinding2020}, the structure of compact stars \cite{oertelEquationsStateSupernovae2017}, and is used in the fitting of parameters for EDFs \cite{benderSelfconsistentMeanfieldModels2003,stoneSkyrmeInteractionFinite2007}. 
Neutron drops trapped in harmonic potentials have also been used to estimate neutron pairing energies in nuclei \cite{smerziNeutronDropsNeutron1997}, to test {\it ab initio} predictions \cite{bogner2011,potter2014,shen2018}, to investigate properties of the tensor force \cite{zhao2020}, as well as to constrain and fit EDF \cite{pudlinerNeutronDropsSkyrme1996,gandolfi2011,maris2013}.
Other examples of simplified models with applications to the nuclear many-body problem include nuclear slabs \cite{bonche1976,rios2011,simenel2014a} and exactly solvable models such as the Lipkin model (see, e.g., Ref.~\cite{severyukhin2006}).

Here, we use a quantum ring, i.e., a one-dimensional system consisting of particles constrained to be around the edge of a circle \cite{viefersQuantumRingsBeginners2004}. 
An advantage of the quantum ring over nuclear matter is that it is not required to be translationally invariant (which for the quantum ring is equivalent to rotational invariance). 
Quantum rings can then be used to investigate rotational symmetry breaking and restoration in mean-field methods. 
Another advantage of the quantum ring lies in its ability to be extended to higher dimensions in order to explore the dependence of the effectiveness of approximate methods on the dimensionality of the system. As an example, a two-dimensional quantum ring, called a spherium, consists of particles constrained to be on the surface of a sphere \cite{loosTwoElectronsHypersphere2009}.
Applications to quantum rings include electron systems \cite{emperador2001,emperador2003,zhu2003,viefersQuantumRingsBeginners2004,fogler2005b,aichinger2006,gylfadottir2006,rasanen2009,manninen2009,loosExactWaveFunctions2012,loos2013,rogers2017,li2021}, electronic and optical properties of semiconductor structures \cite{neillAccuratelyComputingElectronic2021a,hernandezRefractiveIndexChange2022}, molecular rings formed from circular polymers \cite{juddMolecularQuantumRings2020}, 
heavy ion storage rings \cite{steck1996,danared2002,hasseStaticCriteriaExistence2003}, atomic bosons \cite{bargi2010,kaminishi2011,manninen2012,chen2019}, and nuclear systems \cite{brayFermionsLongFiniterange2021}. 

The exact ground-state wave function of two electrons on a ring has been found analytically as an infinite series for all ring radii \cite{zhu2003} and as a series with a finite number of terms for particular radii \cite{loosExactWaveFunctions2012}. In addition, the variational Monte Carlo (VMC)  method was shown to be effective at accounting for correlations in the ground-state of a $N$-fermion quantum ring \cite{gylfadottir2006,brayFermionsLongFiniterange2021}. 
In particular, it was found that the Unbroken-Symmetry Hartree-Fock (USHF) method gave a good approximation to the ground-state in some cases but failed in the case of dense systems with strong short-ranged repulsion as well as attractive systems  \cite{brayFermionsLongFiniterange2021}.
These results lead naturally to the question of the effectiveness of rotational symmetry breaking and restoration in the Hartree-Fock method when applied to the quantum ring, which is the subject of this work. We investigate the effectiveness of the Broken-Symmetry Hartree-Fock (BSHF) and Restored-Symmetry Hartree-Fock (RSHF) methods (see Table~\ref{methods}) under a range of conditions. To compare the accuracy of these methods, we used the VMC method to find a quasi-exact ground-state.

For simplicity, our HF calculations are performed using model interactions rather than through the use of an EDF. 
Consequently, some dynamical correlations that are usually included via the EDF fitting process are not accounted for in the present study. 
Our focus is on the study of static correlations that can be included through symmetry breaking and restoration techniques. 
Although our results are indicative of the efficiency of these techniques, studies with EDF methods should be considered to quantitatively evaluate their efficiency in nuclear systems. 
These, however, are beyond the scope of the current work and will be the purpose of future investigations.

The quantum ring model, interactions and Hamiltonian are introduced in  section \ref{qrm}. 
The many-body methods used in this work are outlined in section \ref{mbm}. 
The results of comparisons between the methods, with a range of  ring radii and strengths of interactions are presented in section \ref{r}. Conclusions are drawn in section \ref{c}.

\begin{table*}
\caption{\label{methods}Many-body methods and properties of the associated wave function.} 
\begin{ruledtabular}
\begin{tabular}{ccp{10cm}}
Method&Acronym&Wave function\\ \hline
Variational Monte Carlo&VMC&Rotationally symmetric parametrized trial wave function.\\
Unbroken Symmetry Hartree-Fock&USHF&Rotationally symmetric Slater determinant of single particle wave functions.\\
Broken Symmetry Hartree-Fock&BSHF& As in USHF without rotational invariance enforced in the wave-function. \\
Restored Symmetry Hartree-Fock&RSHF&Weighted sum of BSHF states using angular momentum projection.\\
\end{tabular}
\end{ruledtabular}
\end{table*}

\section{\label{qrm}Quantum Ring Model}
We use dimensionless distances and energies and set the mass of the fermions on the ring, $m$, and the reduced Planck constant, $\hbar$, to be 1. The quantum ring consists of $N$ fermions constrained to be on a circle of radius $R$. This work does not focus on any effects involving the spin of the particles, so for simplicity, every particle is in the same spin eigenstate in the direction of the axis of the ring. The particles interact via a two-body potential operator, $\hat{V}_{int}(\alpha,\beta)$, which in the position basis is a function of the through-the-ring distance between particles $\alpha$ and $\beta$, $V_{int}(r_{\alpha\beta})$, where
\begin{equation}
r_{\alpha\beta} = 2R\left|\sin\left(\frac{\theta_\alpha-\theta_\beta}{2}\right)\right|
\end{equation}
and $\theta_i$ are the angular coordinates of the particles. 

\begin{figure}
  \begin{center}  
    \includegraphics[width=8.5cm]{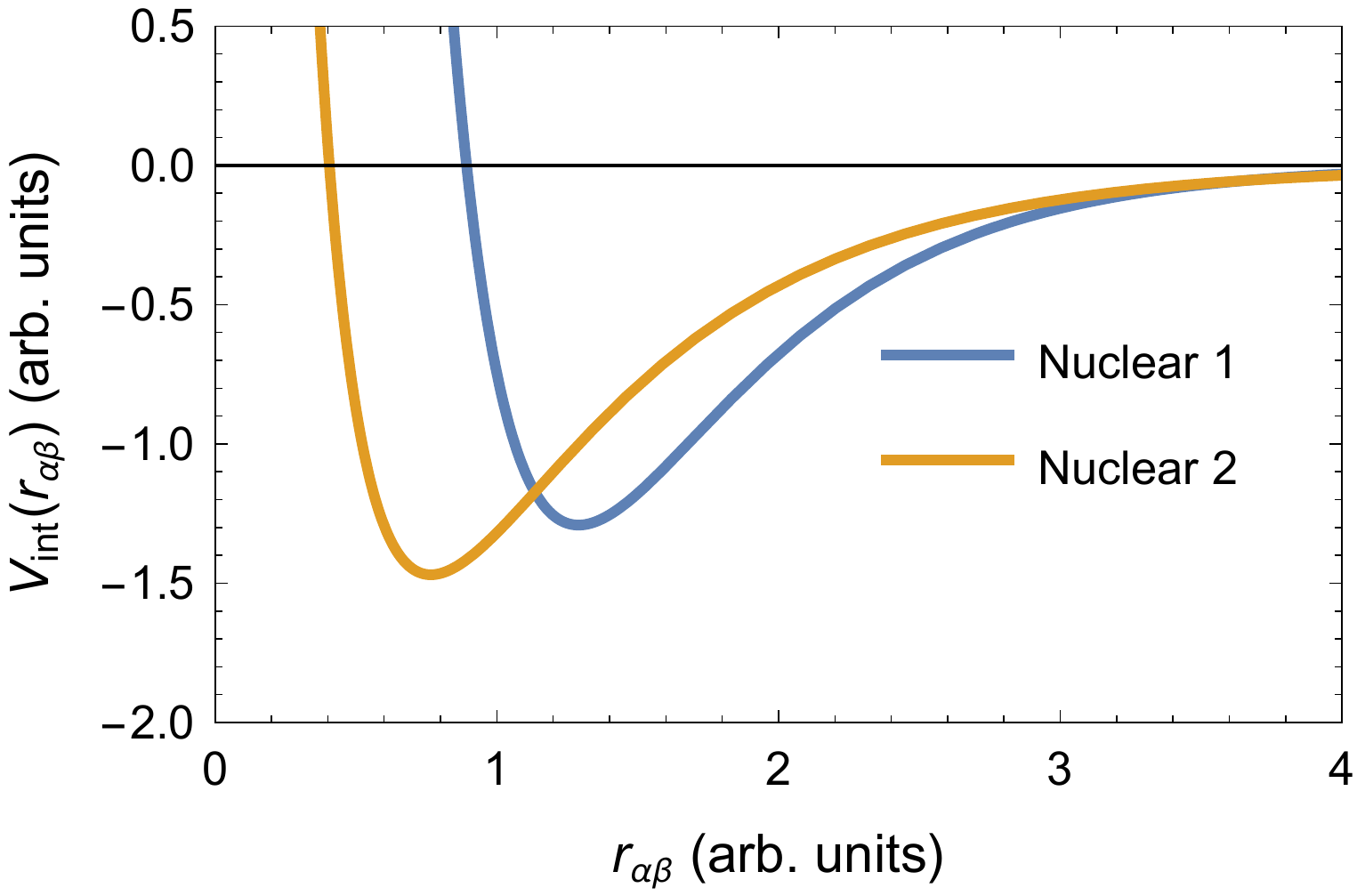}
  \end{center}
\caption{\label{nuclearVints}A comparison between the two `nuclear' potentials used in this work with $V_0=1$.}
\end{figure}

Following Ref.~\cite{brayFermionsLongFiniterange2021}, several interaction potentials are considered. 
The first two interactions are repulsive ($V_{C+}$) and attractive ($V_{C-}$) `Coulomb' potentials,
\begin{equation}
V_{C\pm}(r_{\alpha\beta}) = \frac{V_0}{r_{\alpha\beta}},
\end{equation} 
where $\pm$ refers to the sign of $V_0$. 
Two other interactions, which we refer to as `Nuclear 1' and `Nuclear 2', are used
to model the nuclear force between nucleons (see Fig.~\ref{nuclearVints}), both taken from \cite{brayFermionsLongFiniterange2021}:
\begin{equation}
\label{nuc1general}
V_{N1}(r_{\alpha\beta}) = \frac{V_0}{r_{\alpha\beta}}\left(100e^{-2 r_{\alpha\beta}}-64 e^{-1.5r_{\alpha\beta}}\right),
\end{equation}
\begin{equation}
\label{nuc2general}
V_{N2}(r_{\alpha\beta}) = \frac{V_0}{r_{\alpha\beta}}\left(12e^{-2 r_{\alpha\beta}}-8 e^{-r_{\alpha\beta}}\right),
\end{equation}
where $V_0>0$. 
With the dimensionless units outlined above, the system Hamiltonian in the position basis is 
\begin{equation}
\label{hamiltonian}
H = -\frac{1}{2R^2}\sum_{\alpha=1}^N \frac{\partial^2}{\partial \theta_\alpha^2} + \sum_{\alpha<\beta}^N V_{\text{int}}(r_{\alpha\beta}),
\end{equation}
where 
$V_{\text{int}}$ can be $V_{C+}$, $V_{C-}$, $V_{N1}$ or $V_{N2}$.

\section{\label{mbm}Many-body methods}
\subsection{\label{sb}Hartree-Fock and symmetry breaking}
The Hartree-Fock approximation involves restricting the variational space of many-body wave functions to those which can be expressed as a single Slater determinant of single particle states:
\begin{equation}
|\Psi\rangle = \frac{1}{\sqrt{N!}}
\begin{vmatrix}
    |1:\psi_1\rangle & |2:\psi_1\rangle & \cdots & |N:\psi_1\rangle \\
    |1:\psi_2\rangle & |2:\psi_2\rangle & \cdots & |N:\psi_2\rangle \\
    \vdots & \vdots& \ddots& \vdots\\
    |1:\psi_N\rangle & |2:\psi_N\rangle & \cdots & |N:\psi_N\rangle \\
\end{vmatrix},
\end{equation}
where $|\alpha:\psi_\beta\rangle$ denotes particle $\alpha$ in the single particle state $|\psi_\beta\rangle$. In a given basis of single particle states, $\left\{|\phi_1\rangle,|\phi_2\rangle,\cdots\right\}$, the elements of the one-body density matrix associated with this Slater determinant are 
\begin{equation}
\rho_{ji} = \langle \phi_j|\hat{\rho}|\phi_i\rangle = \sum_{\alpha=1}^N \langle \phi_j|\psi_\alpha\rangle\langle\psi_\alpha|\phi_i\rangle,
\end{equation}
from which we can calculate the Hartree-Fock Hamiltonian as 
\begin{equation}
h[\rho]_{ij} = \langle \phi_i|\hat{h}[\rho]|\phi_j\rangle = \frac{\delta \langle \Psi|\hat{H}|\Psi\rangle}{\delta \rho_{ji}}.
\end{equation}
For a system Hamiltonian $\hat{H}$ comprised of the one-body kinetic energy $\hat{T}$, the one-body external potential $\hat{V}^{\text{ext}}$, and the two-body interaction $\hat{V}^{\text{int}}$,  we get
\begin{equation}
\label{H_in_basis}
h[\rho]_{mn} = T_{mn} + V_{mn}^{\text{ext}} + \sum_{i,j=1}^K \rho_{ji} (V_{imjn}^{\text{int}}-V_{imnj}^{\text{int}}).
\end{equation}

The Hartree-Fock ground-state of a quantum ring which respects the rotational symmetry is a Slater determinant of the $N$-lowest kinetic energy eigenstates of the ring.
These states form a basis of the single-particle Hilbert space and can be written as 
\begin{equation}
\frac{1}{\sqrt{2\pi}},\,\frac{\sin(\theta)}{\sqrt{\pi}},\,\frac{\cos(\theta)}{\sqrt{\pi}},\cdots,\frac{\sin((N-1)\frac{\theta}{2})}{\sqrt{\pi}},\,\frac{\cos((N-1)\frac{\theta}{2})}{\sqrt{\pi}}
\end{equation}
for $N$ odd and
\begin{equation}
\frac{\sin(\frac{\theta}{2})}{\sqrt{\pi}},\,\frac{\cos(\frac{\theta}{2})}{\sqrt{\pi}},\cdots,\,\frac{\sin((N-1)\frac{\theta}{2})}{\sqrt{\pi}},\,\frac{\cos((N-1)\frac{\theta}{2})}{\sqrt{\pi}}
\end{equation}
for $N$ even.
This basis was also used in the other many-body methods. 

To find broken symmetry Hartree-Fock ground-states we used the imaginary time-step method \cite{daviesApplicationImaginaryTime1980}. 
In this method, the energy of the single particle states is lowered through the application of the operator $e^{-d\tau \,\hat{h}}$ which corresponds to a short evolution in imaginary time. 
The time step is chosen as $d\tau=10^{-4}/E_{\text{USHF}}$, where $E_{\text{USHF}}$ is the energy of the unbroken symmetry HF state. 
Gram-Schmidt orthogonalization and normalization of the single-particle states were performed after each iteration.
The unbroken symmetry state was used as an initial state for the imaginary time evolution.
Rotational symmetry was broken with an external potential $\hat{V}^{\text{ext}}$ for the first $10^3$ iterations and progressively removed over the next $10^2$ iterations. 
The specific form of $\hat{V}^{\text{ext}}$ did not impact the final results.
$10^4$ iterations were sufficient to obtain convergence of the BSHF energies.



\subsection{\label{sr}Symmetry restoration with angular momentum projection}
The projector~\cite{sheikhSymmetryRestorationMeanfield2021}
\begin{equation}
\hat{P} = \frac{1}{2\pi}\int_{0}^{2\pi}d\varphi\, e^{-i\varphi(\hat{J}_z-M)}
\end{equation}
is used to project the broken symmetry Hartree-Fock ground-state, $|\Psi_{BS}\rangle$, back onto a state with good angular momentum $M$. 
The true ground-state of a system with only central interaction potentials is expected to have zero orbital angular momentum \cite{mur-loosTwoElectronsHypersphere2009}.
The final total angular momentum should then be entirely determined by the intrinsic spin of the particles. 
As we consider all particles to have spin up, this gives $M=N/2$. 

The energy of the restored symmetry state $\hat{P}|\Psi_{\text{BS}}\rangle$ is given by 
\begin{equation}
E_{\text{SR}} = \frac{\langle\Psi_{\text{BS}}|\hat{P}^\dagger\hat{H}\hat{P}|\Psi_{\text{BS}}\rangle}{\langle\Psi_{\text{BS}}|\hat{P}^\dagger\hat{P}|\Psi_{\text{BS}}\rangle}
=\frac{\langle\Psi_{\text{BS}}|\hat{H}\hat{P}|\Psi_{\text{BS}}\rangle}{\langle\Psi_{\text{BS}}|\hat{P}|\Psi_{\text{BS}}\rangle},
\end{equation} 
which, with some manipulation \cite{CescaHonours}, becomes
\begin{align}
\label{srint}
E_{\text{SR}} =  &\int_{0}^{2\pi} d\varphi\,d\boldsymbol{\theta}\,\frac{\Psi_{\text{BS}}(\boldsymbol{\theta},0)\Psi_{\text{BS}}(\boldsymbol{\theta},\varphi)}{\int_{0}^{2\pi} d\varphi'\, d\boldsymbol{\theta}'\,\Psi_{\text{BS}}(\boldsymbol{\theta}',0)\Psi_{\text{BS}}(\boldsymbol{\theta}',\varphi')}\times\notag\\
& \left(\frac{-1}{2R^2\Psi_{\text{BS}}(\boldsymbol{\theta},\varphi)}\sum_{\alpha=1}^N\frac{\partial^2\Psi_{\text{BS}}(\boldsymbol{\theta},\varphi)}{\partial \theta_\alpha^2} + \sum_{\alpha<\beta}^N V(\theta_\alpha,\theta_\beta)\right),
\end{align}  
where the integration is over the position of each of the particles, $\boldsymbol{\theta} = \{\theta_1,\theta_2,...,\theta_n\}$, and the rotation angle $\varphi$. 
The HF state is the Slater determinant of single particle wave functions,
\begin{equation}
\Psi_{\text{BS}}(\boldsymbol{\theta},0) \propto
\begin{vmatrix}
\psi_1(\theta_1) &   \psi_2(\theta_1) &  \cdots &\psi_n(\theta_1)\\
\psi_1(\theta_2) &   \psi_2(\theta_2)&  \cdots &\psi_n(\theta_2)\\
\vdots & \vdots & \ddots & \vdots\\
\psi_1(\theta_n)&   \psi_2(\theta_n) &  \cdots &\psi_n(\theta_n)
\end{vmatrix}.\label{eq:Slater}
\end{equation}
The rotation by $\varphi$ is defined as $\Psi_{\text{BS}}(\boldsymbol{\theta},\varphi) = \Psi_{\text{BS}}(\theta_1-\varphi, \theta_2-\varphi,...,\theta_N-\varphi,0)$. 
All wave functions are chosen to be real.

To evaluate the integral in equation (\ref{srint}), we defined a sampling distribution 
\begin{equation}
\mathcal{P}_{\text{SR}}(\boldsymbol{\theta},\varphi) = \frac{|\Psi_{\text{BS}}(\boldsymbol{\theta},0)\Psi_{\text{BS}}(\boldsymbol{\theta},\varphi)|}{\int_{0}^{2\pi} d\varphi'\, d\boldsymbol{\theta}'\,\Psi_{\text{BS}}(\boldsymbol{\theta}',0)\Psi_{\text{BS}}(\boldsymbol{\theta}',\varphi')},
\end{equation} 
and the ``local energy'' function~\cite{needs_continuum_2010}
\begin{align}
\mathcal{E}_{\text{SR}}(\boldsymbol{\theta},\varphi)&=\text{sign}\left(\Psi_{\text{BS}}(\boldsymbol{\theta},0)\Psi_{\text{BS}}(\boldsymbol{\theta},\varphi)\right)\times\nonumber\\
&\left(\frac{-1}{2R^2\Psi_{\text{BS}}(\boldsymbol{\theta},\varphi)}\sum_{\alpha=1}^N\frac{\partial^2\Psi_{\text{BS}}(\boldsymbol{\theta},\varphi)}{\partial \theta_\alpha^2} + \sum_{\alpha<\beta}^N V(\theta_\alpha,\theta_\beta)\right),
\end{align}
then used Monte Carlo integration with the Metropolis algorithm \cite{1953JChemPhysMetropolis} (see next subsection) to evaluate 
\begin{equation}
E_{\text{SR}} =  \int_{0}^{2\pi} d\varphi\,d\boldsymbol{\theta}\, \mathcal{P}_{\text{SR}}(\boldsymbol{\theta},\varphi)\mathcal{E}_{\text{SR}}(\boldsymbol{\theta},\varphi).
\end{equation}

\subsection{\label{vmc}Variational Monte Carlo}
In the variational Monte Carlo method, we consider a parametrized trial wave function, $\Psi_{\text{VMC}}(\boldsymbol{\theta},\boldsymbol{a})$ which is more general than the Slater determinant used in the Hartree-Fock method. The aim of the method is to minimise the energy of the trial wave function with respect to its parameters, $\boldsymbol{a}$, to get closer to the system's true ground-state. The energy of a trial wave function is given by 
\begin{equation}
\label{avenergy}
E_{\text{VMC}}(\boldsymbol{a}) = \frac{\int d\boldsymbol{\theta}\,\Psi_{\text{VMC}}^*(\boldsymbol{\theta},\boldsymbol{a})\, H\,\,\Psi_{\text{VMC}}(\boldsymbol{\theta},\boldsymbol{a})}{\int d\boldsymbol{\theta}\,\Psi_{\text{VMC}}^*(\boldsymbol{\theta},\boldsymbol{a}) \Psi_{\text{VMC}}(\boldsymbol{\theta},\boldsymbol{a})},
\end{equation}
which can be written as 
\begin{equation}
E_{\text{VMC}}(\boldsymbol{a}) =  \int_{0}^{2\pi} d\boldsymbol{\theta}\, \mathcal{P}_{\text{VMC}}(\boldsymbol{\theta},\boldsymbol{a})\mathcal{E}_{\text{VMC}}(\boldsymbol{\theta},\boldsymbol{a}),
\end{equation}
where the sampling distribution is 
\begin{equation}
 \mathcal{P}_{\text{VMC}}(\boldsymbol{\theta},\boldsymbol{a}) = \frac{|\Psi_{\text{VMC}}(\boldsymbol{\theta},\boldsymbol{a})|^2}{\int_{0}^{2\pi} d\boldsymbol{\theta}'\,|\Psi_{\text{VMC}}(\boldsymbol{\theta}',\boldsymbol{a})|^2},
\end{equation} 
and the ``local energy'' function
\begin{equation}
\mathcal{E}_{\text{VMC}}(\boldsymbol{\theta},\boldsymbol{a}) = \frac{-1}{2R^2\Psi_{\text{VMC}}}\sum_{\alpha=1}^N\frac{\partial^2\Psi_{\text{VMC}}}{\partial \theta_\alpha^2} + \sum_{\alpha<\beta}^N V(\theta_\alpha,\theta_\beta).
\end{equation}
The integrals are too complicated to solve analytically, so the method uses Monte Carlo integration with the Metropolis algorithm~\cite{1953JChemPhysMetropolis}. Samples of the integration space, $\left\{\boldsymbol{\theta_1},\boldsymbol{\theta_2},\boldsymbol{\theta_3}...\right\}$ are chosen sequentially from the probability distribution $\mathcal{P}_{\text{VMC}}$ and an average of the function $\mathcal{E}_{\text{VMC}}$ of each of these samples is taken.

The statistical uncertainty in the wave function's energy was calculated using the blocking method \cite{flyvbjergErrorEstimatesAverages1989} since the Metropolis algorithm results in correlated samples of the integration space. These uncertainties only reflect the uncertainty in the energy of the trial wave function and are not a measure of how close the trial wave function energy is to the exact ground-state energy, which is difficult to determine for most systems.

In this work, we used a Jastrow trial wave function \cite{jastrowManyBodyProblemStrong1955}, i.e., a Slater determinant of single particle wave functions multiplied by a Jastrow correlation factor which consists of the product of strictly positive functions of the inter-particle distances, $r_{\alpha\beta}$, 
\begin{align}
\Psi_{\text{VMC}}(\theta_1,&\theta_2,...,\theta_n,\boldsymbol{a}) = \nonumber\\
&\begin{vmatrix}
    \phi_1(\theta_1) & \phi_1(\theta_2) & \cdots & \phi_1(\theta_n) \\
    \phi_2(\theta_1) & \phi_2(\theta_2) & \cdots & \phi_2(\theta_n) \\
    \vdots & \vdots& \ddots& \vdots\\
    \phi_n(\theta_1) & \phi_n(\theta_2) & \cdots & \phi_n(\theta_n) \\
\end{vmatrix}\prod_{\alpha<\beta}^N f(r_{\alpha\beta},\boldsymbol{a}).
\end{align}
The single particle wave functions of the Slater determinant were chosen to be the kinetic energy eigenstates of the system which were also the Hartree-Fock ground-state without broken rotational symmetry. The correlation factor was chosen to be an exponential of a polynomial of the inter-particle distances 
\begin{equation}
\prod_{\alpha<\beta}^N f(r_{\alpha\beta},\boldsymbol{a}) = \exp\left(\sum_{\alpha<\beta}^N \sum_{k=1}^4 a_k r_{\alpha\beta}^k\right).
\end{equation}
The parameters of the trial wave function were optimised using the DIRECT (DIviding RECTangles) algorithm \cite{jonesLipschitzianOptimizationLipschitz1993} included in the NLopt non-linear optimisation package \cite{steven_g_johnson_nlopt_nodate}.
Table \ref{compwithloos} gives the VMC results for two electrons on rings of differing radii together  with the exact analytical results found by Loos and Gill \cite{loosExactWaveFunctions2012} as well as the Hartree-Fock ground-state with rotational symmetry. 

\begin{table*}
\caption{\label{compwithloos}Ground-state energies of a quantum ring with $N=2$ particles and a repulsive `Coulomb' interaction of strength $V_0=1$. All quantities are dimensionless.}
\begin{ruledtabular}
\begin{tabular}{cccc}
Ring radius& Analytical \cite{loosExactWaveFunctions2012}& VMC &Hartree Fock\\
 \hline
$1/2$ & $9/4 = 2.2500$ & $2.2503 \pm 0.0008$ & $2.2732$\\
$\sqrt{3/2}$ & $2/3 = 0.66667$ & $0.66661 \pm 0.00016$ & $0.68646$\\
$\sqrt{\frac{3}{4}(7+\sqrt{33})}$ & $\frac{25}{96}(7-\sqrt{33}) = 0.32694$ & $0.32695 \pm 0.00006$ & $0.34352$\\
$\sqrt{23/2}$ & $9/46 = 0.19565$ & $0.19562 \pm 0.00003$ & $0.20947$\\
\end{tabular}
\end{ruledtabular}
\end{table*}


\section{\label{r}Results}

Ground-state energies have been computed using each of the many-body methods  
for various ring radii, particle numbers, and interaction types and strengths.
For each system, the correlation energy is defined as the difference in the ground-state energy between the USHF method (which includes no correlations between the particles) and the VMC method. Then for the BSHF and RSHF results we calculated the percentage of correlation energy that they account for,
\begin{equation}
\frac{E_{\text{USHF}}-E_{\text{BSHF}}}{E_{\text{USHF}}-E_{\text{VMC}}} \text{     or     } \frac{E_{\text{USHF}}-E_{\text{RSHF}}}{E_{\text{USHF}}-E_{\text{VMC}}},\label{eq:correlE}
\end{equation}
as a measure of each method's effectiveness.

The linear particle density is defined from the BSHF single particle wave functions $\Psi_i(\theta)$ [see Eq.~(\ref{eq:Slater})] as 
\begin{equation}
\rho_{\mathrm{lin.}}(R\theta)=\frac{1}{R}\sum_{i=1}^N\psi_i(\theta)^2.
\label{eq:linear_density}
\end{equation}
Similarly, the angular density is defined as 
\begin{equation}
\rho_{\mathrm{ang.}}(\theta)=\sum_{i=1}^N\psi_i(\theta)^2.
\label{eq:angular_density}
\end{equation}
The normalisation $\int_0^{2\pi}d\theta\,\psi_i(\theta)^2=1$ leads to $\int_0^{2\pi R} d(R\theta) \,\rho_{\mathrm{lin.}}(R\theta)=\int_0^{2\pi} d\theta \,\rho_{\mathrm{ang.}}(\theta)=N$.

\subsection{Varying ring radius}
\subsubsection{Attractive `Coulomb'}
\begin{figure}
  \begin{center}  
    \includegraphics[width=8.5cm]{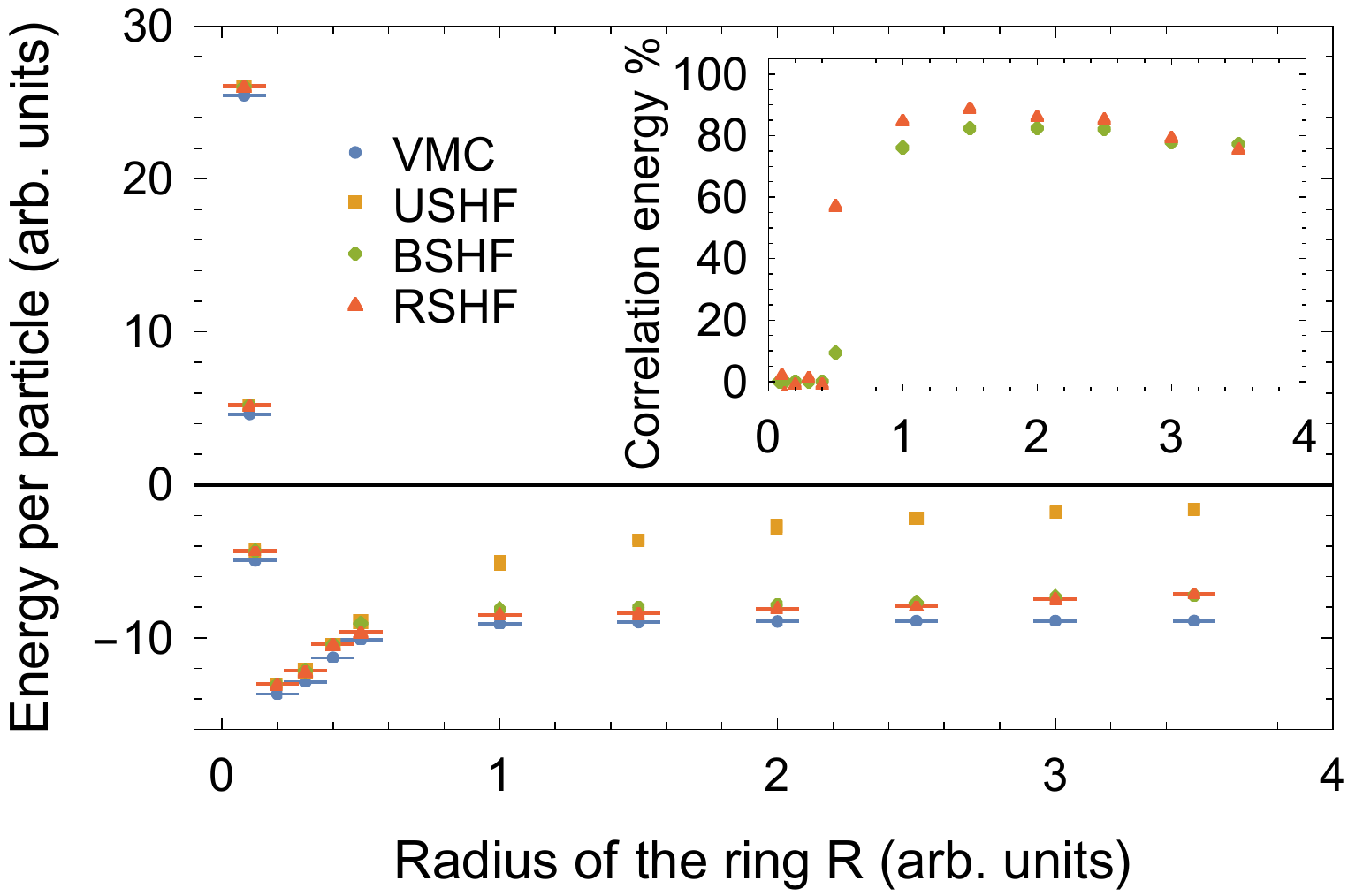}
  \end{center}
\caption{\label{RvaryCminus}Energy per particle for a quantum ring with attractive `Coulomb' interactions of strength $V_0=5$, $N=4$ particles, and varying radii of the ring, $R$. Inset shows the percentage of correlation energy accounted for by the BSHF and RSHF methods [see Eq.~(\ref{eq:correlE})]. The horizontal axis of the inset is the same as for the main figure.}
\end{figure}
\begin{figure}
  \begin{center}  
    \includegraphics[width=8.5cm]{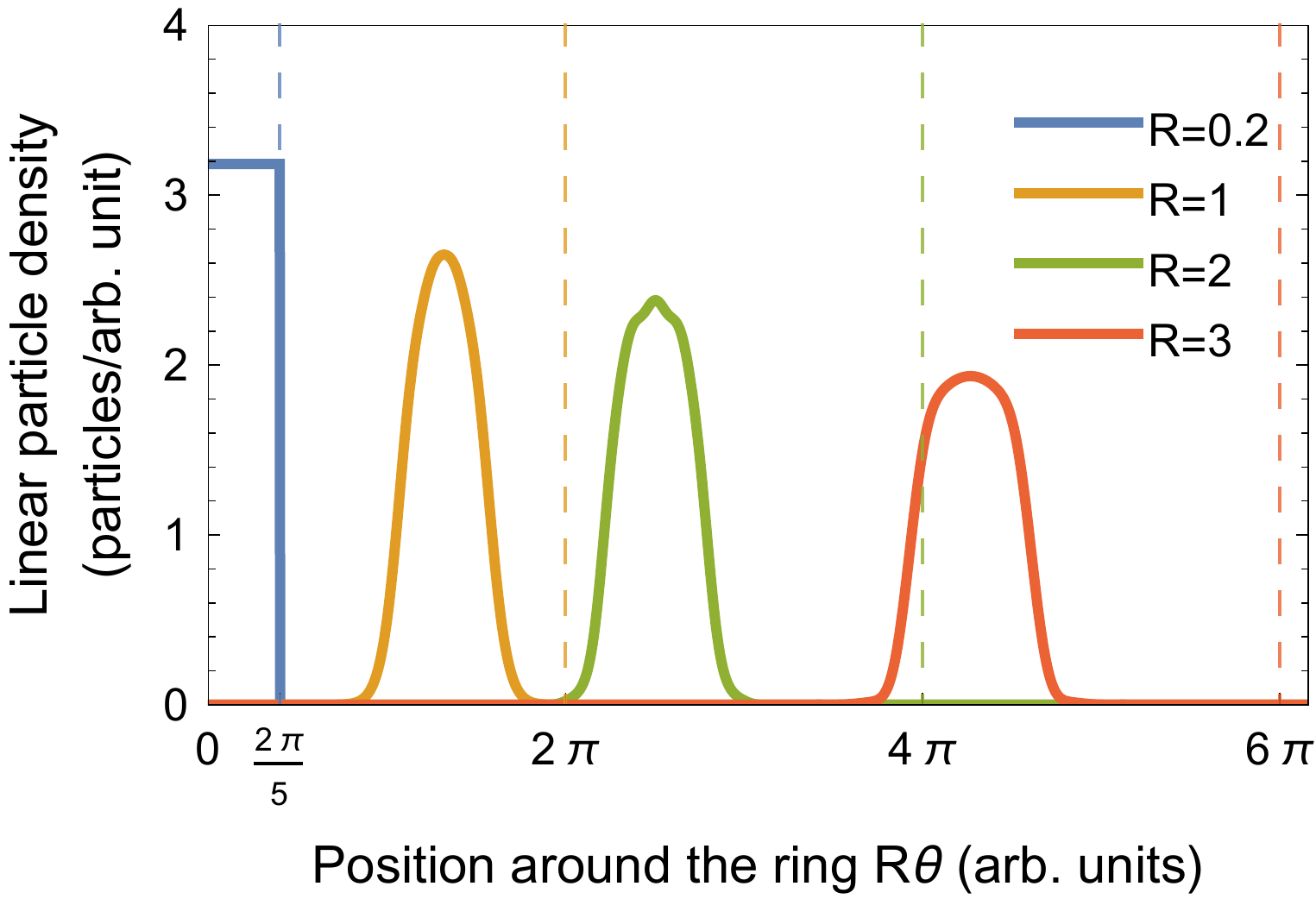}
  \end{center}
\caption{\label{RvaryCminusdensity}Linear particle density of the BSHF ground-state [see Eq.~(\ref{eq:linear_density})] with the attractive `Coulomb' interaction of strength $V_0=5$ and $N=4$ particles. The vertical lines indicate the maximum value the position can take on each ring (i.e., $2\pi R$).}
\end{figure}
Figure~\ref{RvaryCminus} shows the effect of varying the ring radius on the energy per particle for the attractive `Coulomb' interaction. Each method gives a large positive energy at small radii ($R<0.2$) which decreases to a negative minimum before increasing to a constant value which is 0 for the USHF method and a negative value for the other three methods. Given that the attractive `Coulomb' potential is strictly negative, the large positive energy at small radii is entirely due to the kinetic energy of the particles which increases due to Pauli repulsion as the ring is made smaller and the particles are forced to overlap more. The USHF energy converges to 0 at large radii since it entirely excludes correlations in the positions of the particles and as a result the particles become further and further apart on average as the radius increases. In the other three methods, the particles are able to correlate their positions on the ring, forming a self-bound system which has a ground-state wave function which is no longer dependent on the radius of the ring. 

Spatial correlations can be observed on the BSHF particle density plotted  in Fig.~\ref{RvaryCminusdensity}. 
for rings of different sizes. On the smallest ring, no symmetry breaking occurs so the  density is constant. 
We see that, for rings with $R\gtrsim1$, 
the particles {\it `bunch together'}, forming a self-bound system whose size and shape becomes mostly independent of the size of the ring for large radii.

Returning to Fig.~\ref{RvaryCminus}, both the BSHF and RSHF methods bring little to no improvement to the ground-state energy at $R<0.5$, since the USHF already gives a good approximation to the exact ground-state energy and no symmetry breaking occurred. At larger radii, the rotational symmetry breaking becomes more significant since the USHF method is unable to model particles bunching on one region of the ring. 
The inset of Fig.~\ref{RvaryCminus} shows that the BSHF method accounts for about $\approx80\%$ of the correlation energy at $R\gtrsim1$ [see Eq.~(\ref{eq:correlE})], while symmetry restoration (RSHF) brings little improvement to the BSHF result.

\subsubsection{`Nuclear 1'}
\begin{figure}
  \begin{center}  
    \includegraphics[width=8.5cm]{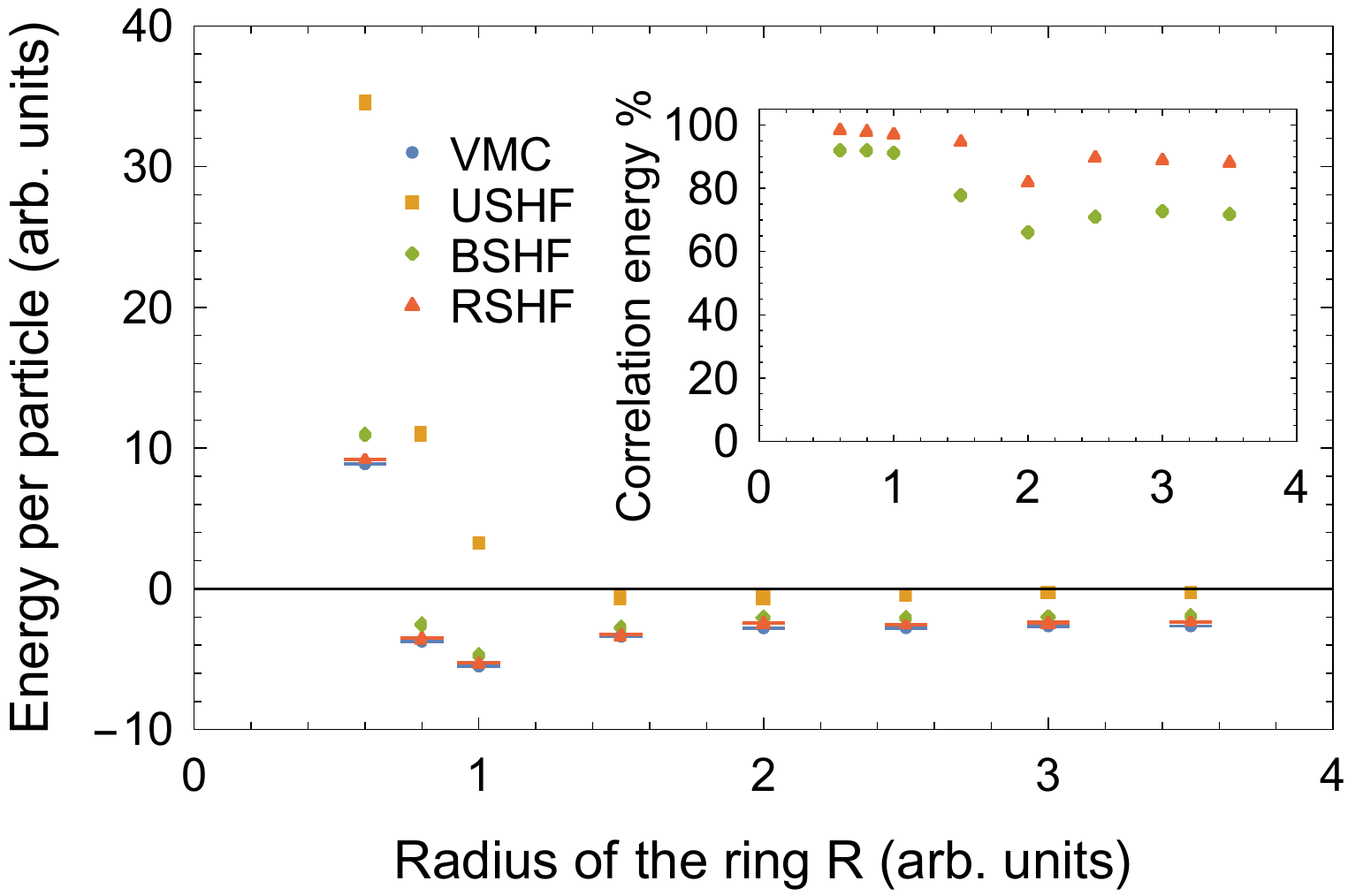} 
  \end{center}
\caption{\label{RvaryN1}Same as Fig.~\ref{RvaryCminus} for the `Nuclear 1' interaction ($V_0=5$, $N=4$).}
\end{figure}
\begin{figure}
  \begin{center}  
    \includegraphics[width=8.5cm]{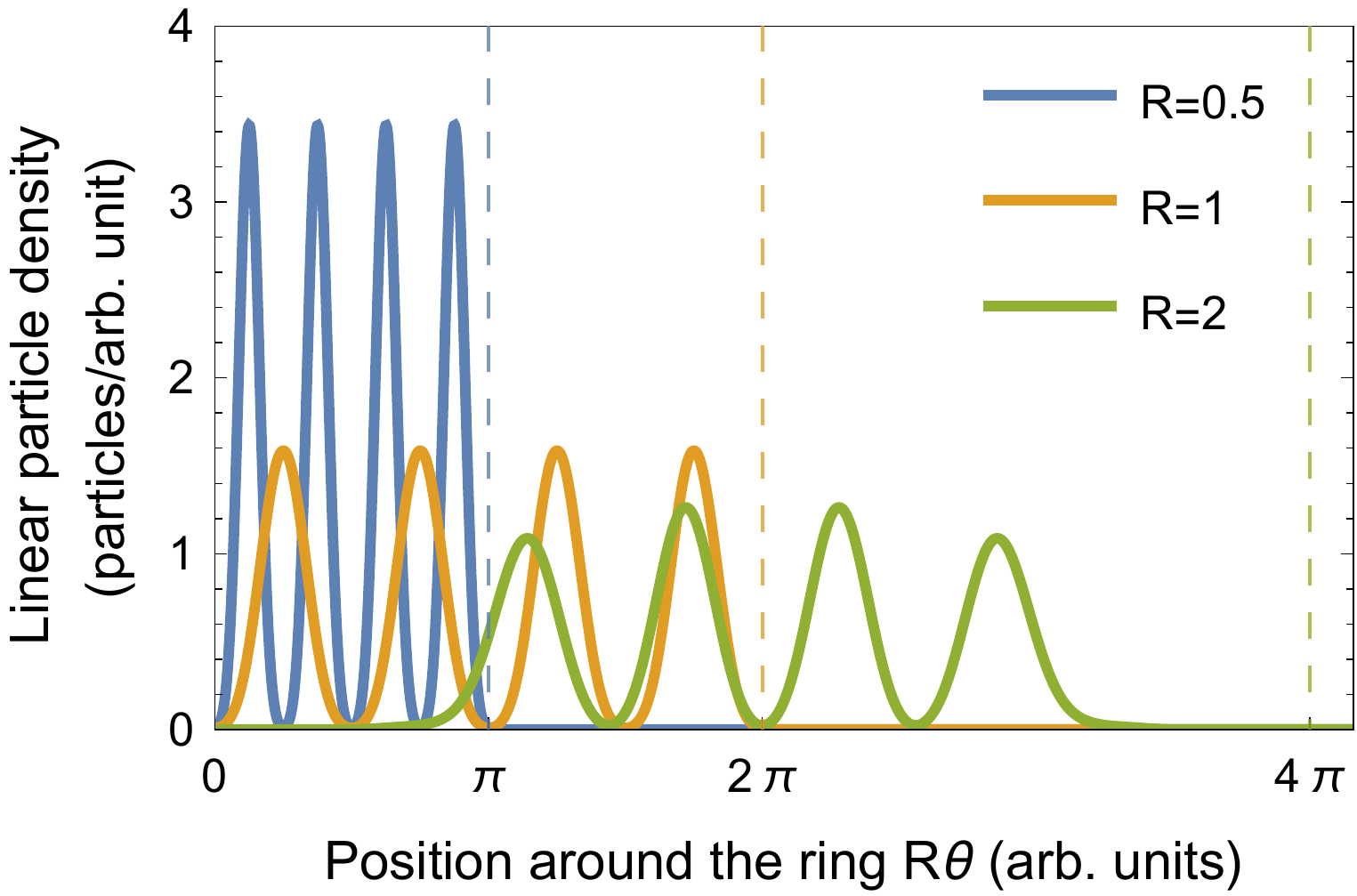}
  \end{center}
\caption{\label{RvaryN1density}Same as Fig.~\ref{RvaryCminusdensity} for the `Nuclear 1' interaction ($V_0=5$, $N=4$).}
\end{figure}

Figure~\ref{RvaryN1} shows the energy per particle for the `Nuclear 1' interaction. In a similar way to the attractive `Coulomb' case, each method gives a large positive ground-state energy at small radii before decreasing to some minimum negative energy per particle then increasing again. Again, the USHF method goes to 0 as $R\rightarrow\infty$ since the particles are not able to correlate their positions while the other three methods converge to constant negative values as they form a self-bound system due to the attractive component of the `Nuclear 1' interaction. Because of its hard-core repulsion, the position of the minimum for this interaction is at a noticeably larger radius than the attractive `Coulomb' case. 

As shown in the inset of Fig.~\ref{RvaryN1}, the BSHF method is able to account for a large portion of the correlation energy, particularly at small radii. 
Symmetry restoration (RSHF) performs better for this interaction, accounting for most of the remaining correlation energy at small radii, making it almost as good as the VMC method. Unlike the attractive `Coulomb' case, the RSHF method also performs well at larger radii, improving the ground-state by around 20\% of the correlation energy.

The density plot in Fig.~\ref{RvaryN1density} demonstrates two different types of correlation between particles induced by the `Nuclear 1' interaction. At lower radii, the particles localise their positions by forming four equally sized peaks spaced evenly around the ring. We call this localisation {\it`spreading out'} since it results in a larger average distance between the particles (compared to the unbroken symmetry state) in order to minimise the hard-core repulsion of the `Nuclear 1' potential. As the radius increases to the $R=2$ case, the particles localise to one region of the ring (bunching) in a similar way to the attractive `Coulomb' case, due to the attractive component of `Nuclear 1'. Unlike the attractive `Coulomb' case, however, the self-bound system does not exhibit a single peak. Instead it consists of four evenly spaced peaks within the self-bound system due to the strong hard-core repulsion. 
The improvement of the RSHF method at larger radii could be a result of this spreading out that breaks the uniformity of the internal density, while the attractive `Coulomb' case, in which spreading out does not occur, is not improved significantly by symmetry restoration.

\subsubsection{`Nuclear 2'}
\begin{figure}
  \begin{center}  
    \includegraphics[width=8.5cm]{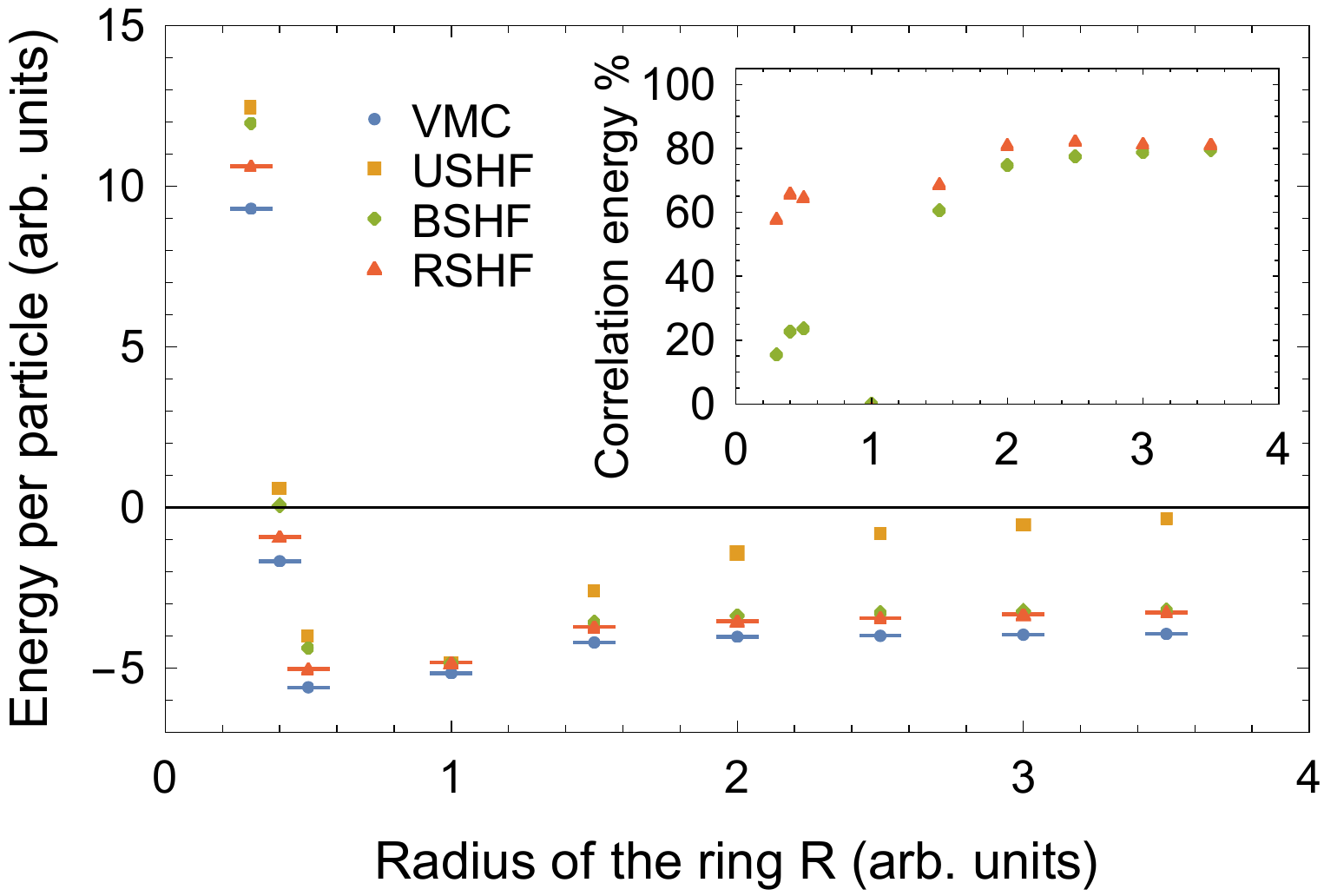}
  \end{center}
\caption{\label{RvaryN2}Same as Fig.~\ref{RvaryCminus} for the `Nuclear 2' interaction ($V_0=5$, $N=4$).}
\end{figure}
\begin{figure}
  \begin{center}  
    \includegraphics[width=8.5cm]{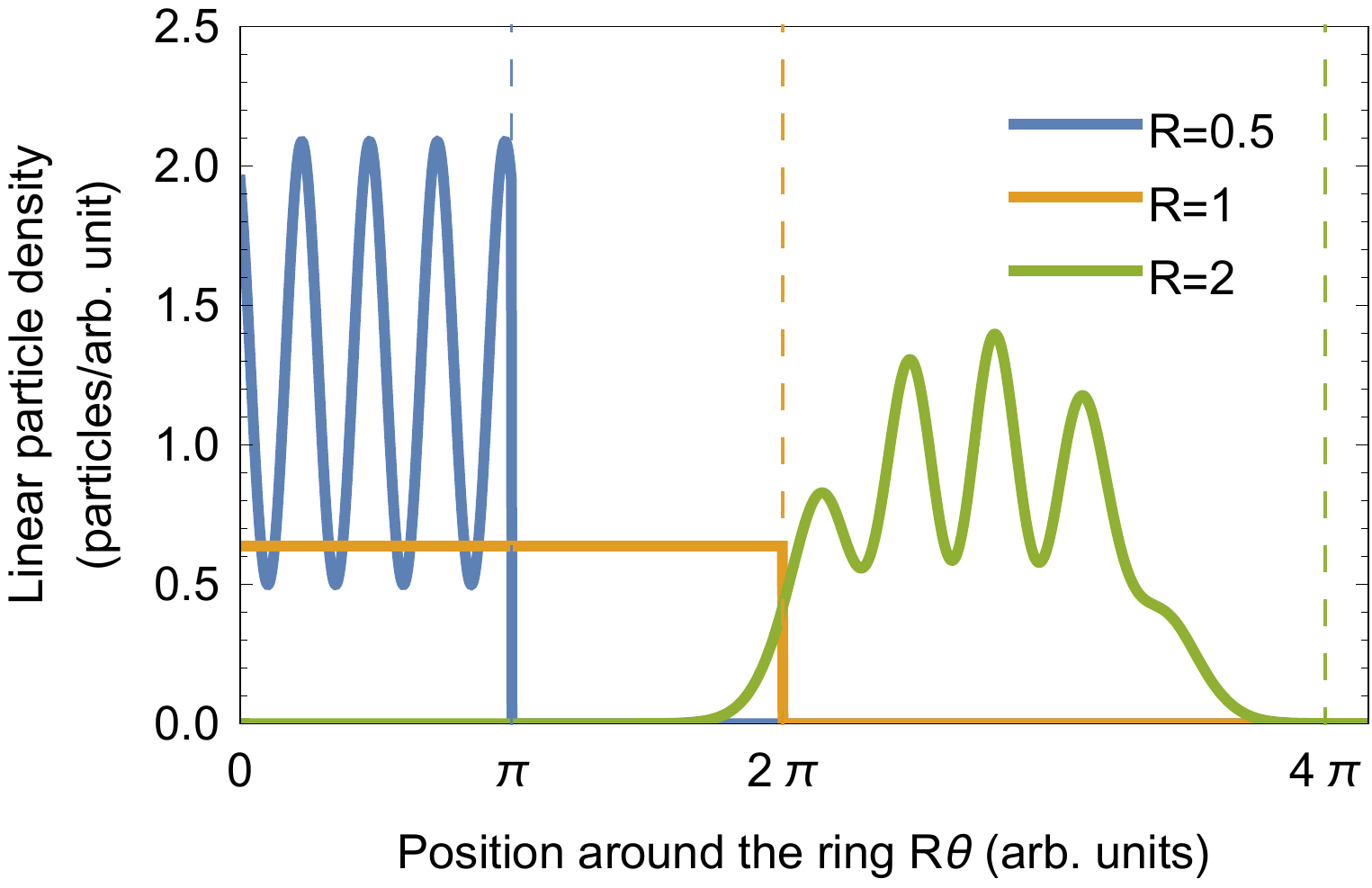}
  \end{center}
\caption{\label{RvaryN2density}Same as Fig.~\ref{RvaryCminusdensity} for the `Nuclear 2' interaction ($V_0=5$, $N=4$).}
\end{figure}

Figure~\ref{RvaryN2} demonstrates the radius dependence of the ground-state energy for the `Nuclear 2' interaction. The results are similar to the `Nuclear 1' case with a few changes. Most notably, the USHF result performs much better at smaller radii with this interaction than for `Nuclear 1' due to the smaller hard-core repulsion leading to less correlation in the exact ground-state. As a result, the BSHF and RSHF give less of an improvement for these ring sizes than they did for `Nuclear 1', especially for $R=1$ where the USHF energy almost equals the exact ground-state. For smaller radii, when the particles mainly repel each other, symmetry restoration (RSHF) accounts for about $\approx60\%$ of the correlation energy (see inset of Fig.~\ref{RvaryN2}) and is then able to improve significantly on the BSHF method ($\approx20\%$ of the correlation energy) while for larger radii both methods give similar energies, accounting for $\approx80\%$ of the correlation energy at $R\gtrsim2$. Another difference when compared to `Nuclear 1' is that the minimum in the energy has shifted to smaller radii which is a result of the weaker hard-core repulsion of the `Nuclear 2' potential.

Figure~\ref{RvaryN2density} shows the particle density for the `Nuclear 2' interaction on rings with varying radii. In a similar way to the `Nuclear 1' density, there is `spreading out' on the ring of radius $R=0.5$ and `bunching with spreading out' on the ring of radius $R=2$. At $R=1$, however, no symmetry breaking occurs resulting in a flat density. 
This is consistent with the ground-state energy (see Fig.~\ref{RvaryN2}) which is predicted to be the same for the BSHF, RSHF, and USHF methods  at $R=1$. This is likely due to the effects of the attractive and repulsive components of the `Nuclear 2' interaction cancelling each other out for this particular ring radius.

\subsection{Varying interaction potential strength}
\subsubsection{Repulsive `Coulomb'}
\begin{figure}
  \begin{center}  
    \includegraphics[width=8.5cm]{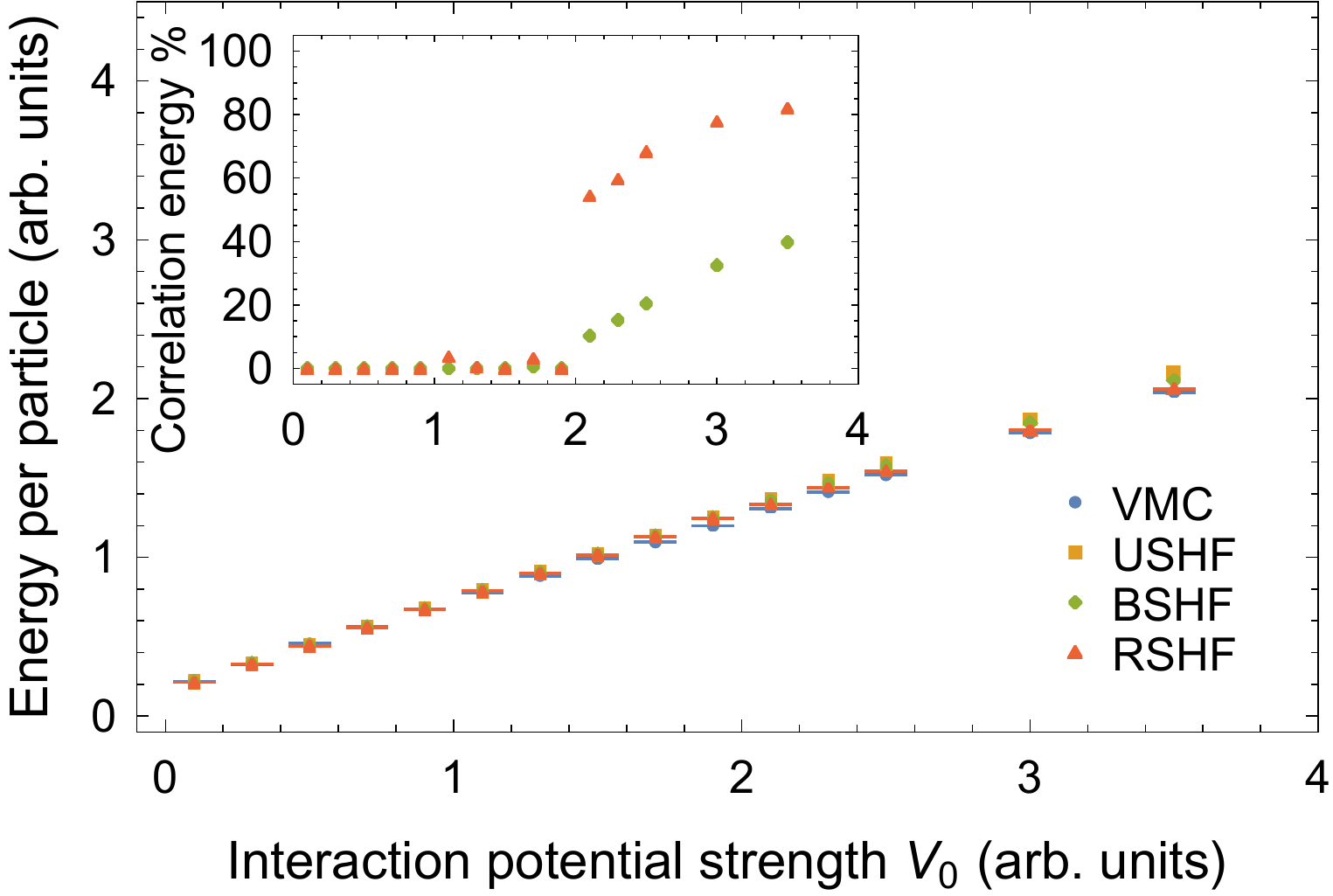}
  \end{center}
\caption{\label{V0varyCplus}Energy per particle for a quantum ring with repulsive `Coulomb' interactions, radius $R=2$, $N=4$ particles and varying strength of the interaction potential, $V_0$. Inset shows the percentage of correlation energy accounted for by the BSHF and RSHF methods [see Eq.~(\ref{eq:correlE})]. The horizontal axis of the inset is the same as for the main figure.}
\end{figure}
Figure~\ref{V0varyCplus} shows the $V_0$ dependence of the energy for the repulsive `Coulomb' interaction. The USHF has a linear dependence on $V_0$ since its kinetic energy per particle is constant and its potential energy is directly proportional to $V_0$. In the other three methods at larger $V_0$ values, the potential energy contribution is slightly reduced by localising the particle positions. 
As a result of the Heisenberg uncertainty principle, a wave function with more localisation in position typically has a higher kinetic energy than a flatter wave function, so it is only for systems with stronger interaction potentials that the localisation occurs and the rotational symmetry is broken. The transition between `not spreading out' and `spreading out' is indicated by the energy per particle falling below the USHF energy. For the BSHF and RSHF methods this transition occurs around $V_0=1.7$. Once the symmetry breaking has occurred in the BSHF method, both the BSHF and RSHF methods increase their fraction of correlation energy with the potential strength, with the RSHF bringing significant improvement to the BSHF ground-state energy (see inset of Fig.~\ref{V0varyCplus}).

\subsubsection{`Nuclear 1'}
\begin{figure}
  \begin{center}  
    \includegraphics[width=8.5cm]{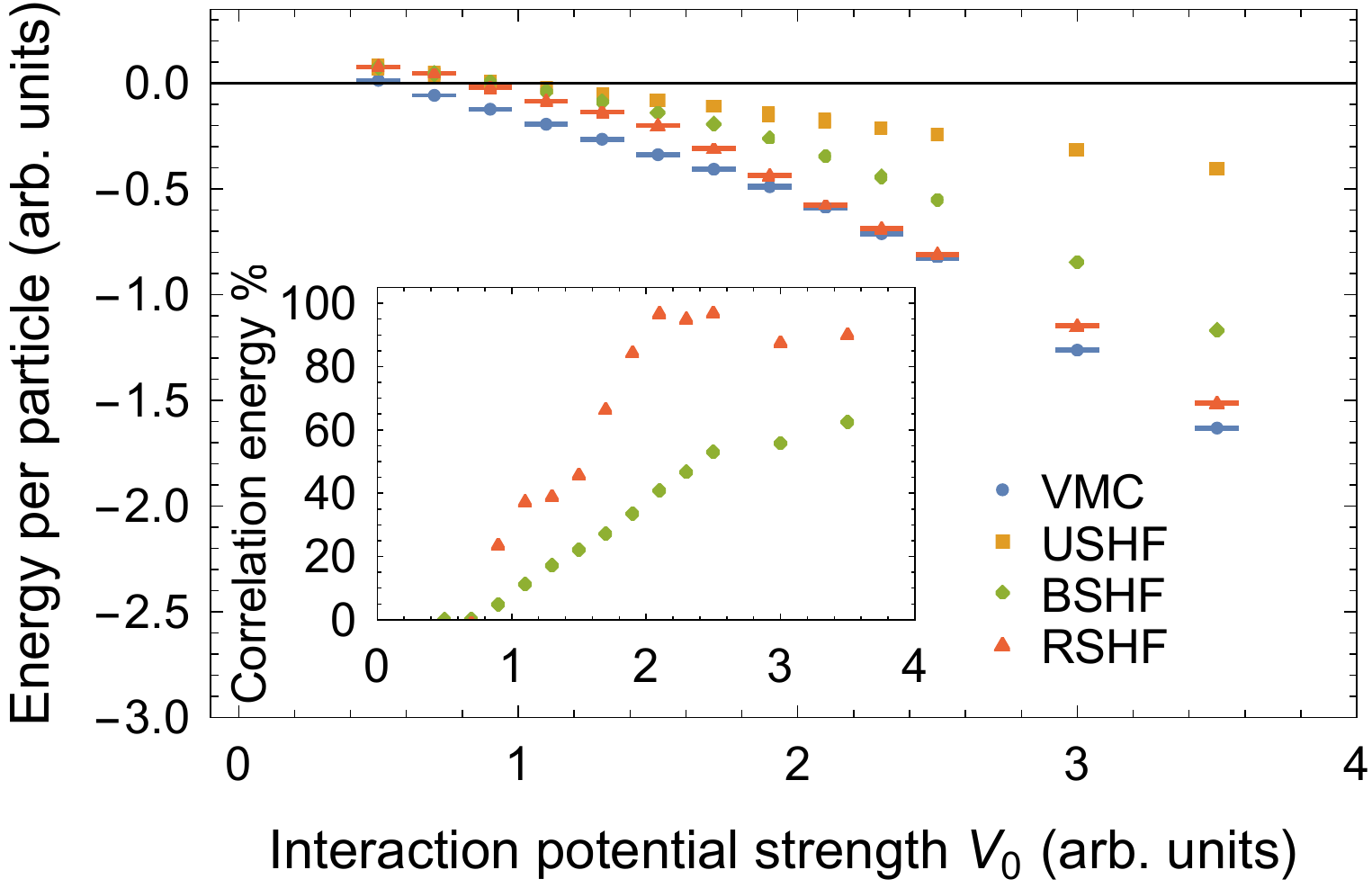}
  \end{center}
\caption{\label{V0varyN1}Same as Fig.~\ref{V0varyCplus} for the `Nuclear 1' interaction ($R=2$, $N=4$).}
\end{figure}
\begin{figure}
  \begin{center}  
    \includegraphics[width=8.5cm]{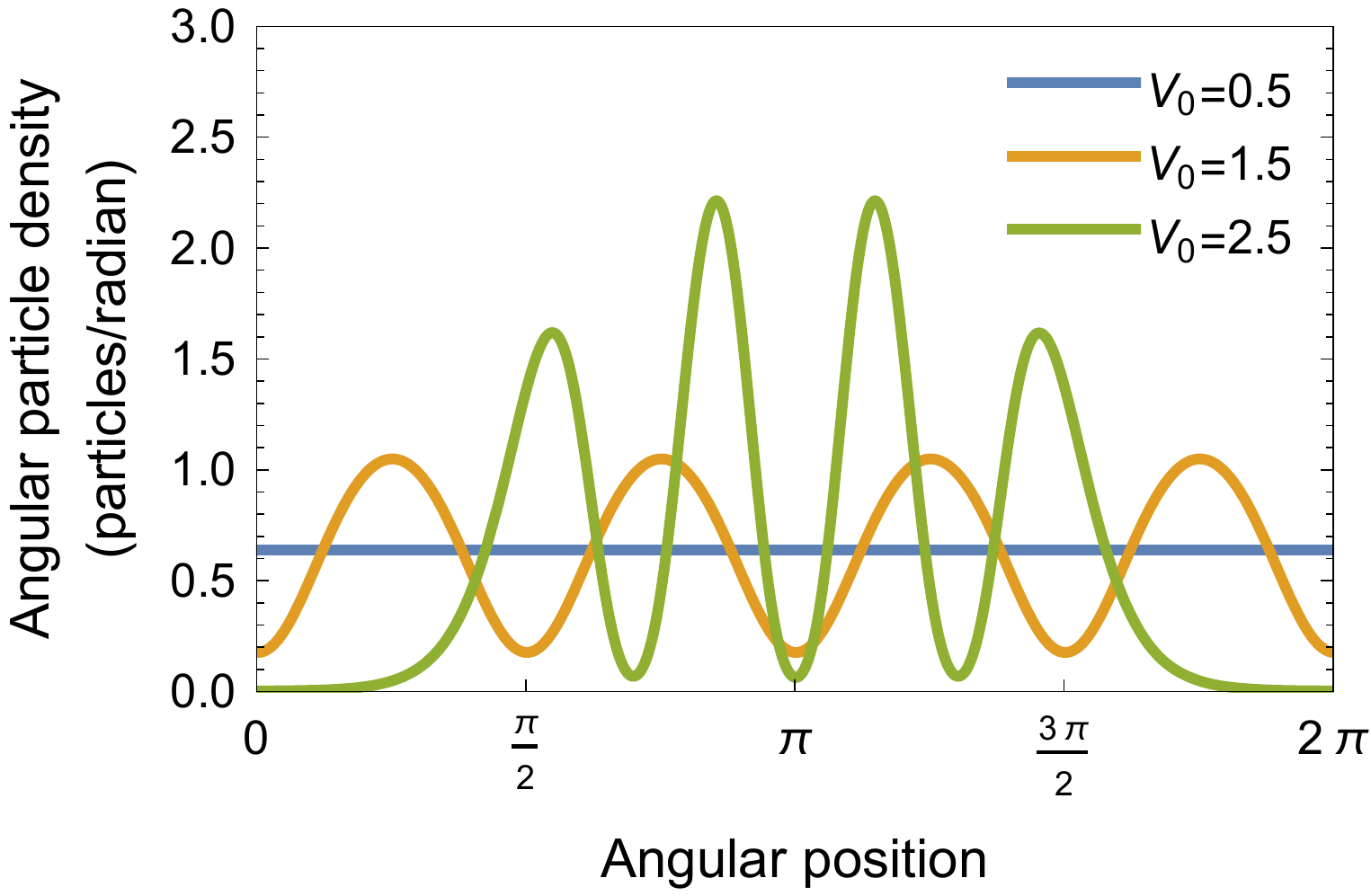}
  \end{center}
\caption{\label{V0varyN1density}Angular particle densities of the BSHF ground-state [see Eq.~(\ref{eq:angular_density})] with the `Nuclear 1' interaction ($R=2$, $N=4$).}
\end{figure}

Figure~\ref{V0varyN1} shows the $V_0$ dependence for the `Nuclear 1' interaction. Once again the USHF method has a linear dependence for the same reason as the repulsive `Coulomb' case. For small $V_0$, the energy per particle is positive due to the dominance of kinetic energy. As $V_0$ increases, the attractive component of `Nuclear 1' potential energy takes over and the energy becomes negative. The other methods also show a steady decrease in energy. 

Figure~\ref{V0varyN1density} shows the angular particle density for the BSHF result with the `Nuclear 1' interaction and different values for $V_0$. 
Three phases are observed: one with no symmetry breaking at $V_0\lesssim0.8$, one with spreading out only for $0.8\lesssim V_0\lesssim1.6$, and one with both spreading out and bunching at $V_0\gtrsim1.6$. 
Returning to Fig.~\ref{V0varyN1}, we see that the performance of the BSHF and RSHF methods, shown in the inset, changes significantly between phases. Clearly, both methods perform the worst when no symmetry is broken ($V_0\lesssim0.8$). Then the performance of both methods increase when the particles are only spreading out, before beginning to flatten out when both spreading out and bunching occurs, for $V_0\gtrsim1.6$. The RSHF method seems especially sensitive to the phase transitions with sharp jumps in performance at the boundaries of the phases.

\subsubsection{`Nuclear 2'}
\begin{figure}
  \begin{center}  
    \includegraphics[width=8.5cm]{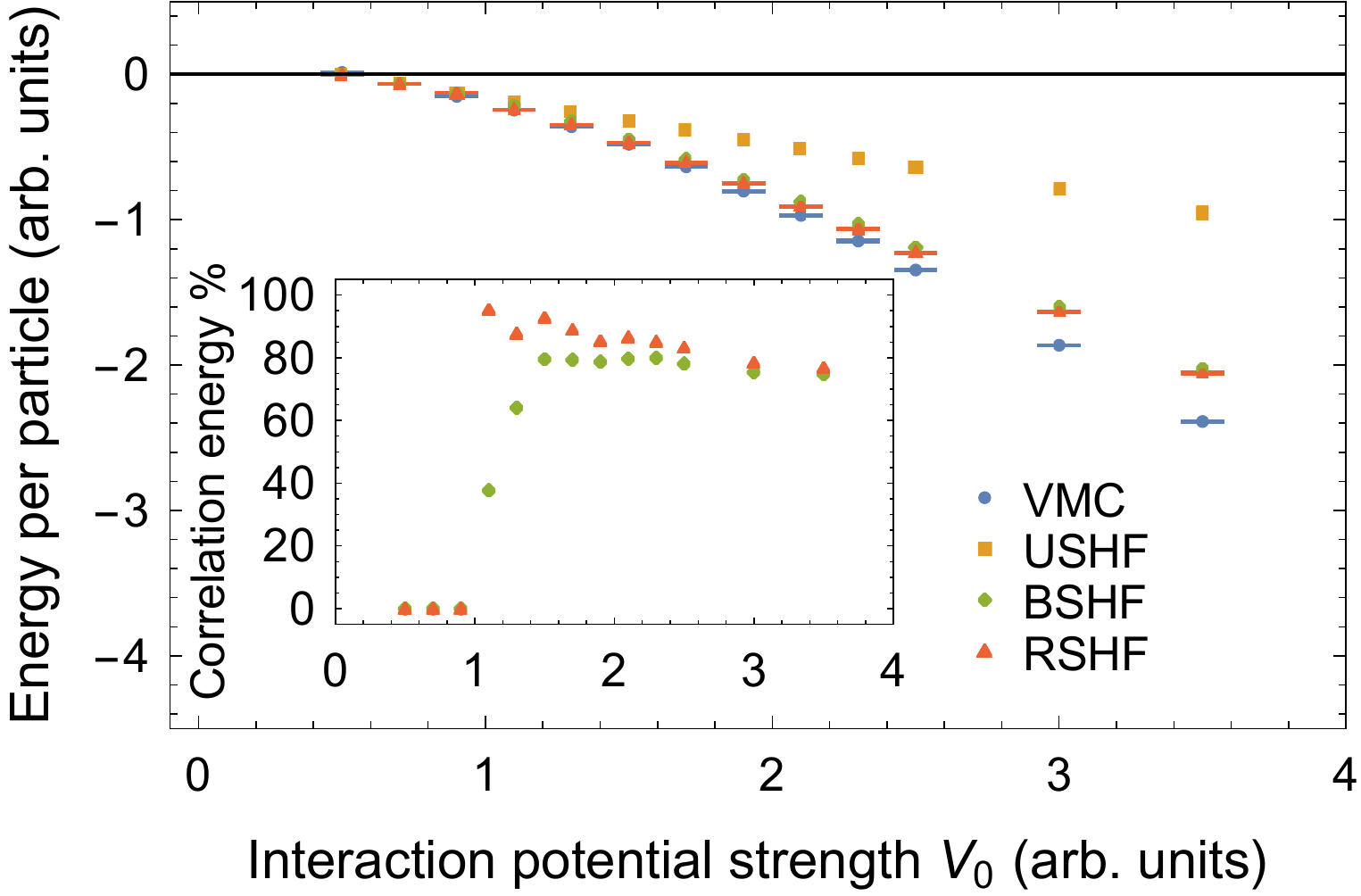}
  \end{center}
\caption{\label{V0varyN2}Same as Fig.~\ref{V0varyCplus} for the `Nuclear 2' interaction ($R=2$, $N=4$).}
\end{figure}
\begin{figure}
  \begin{center}  
    \includegraphics[width=8.5cm]{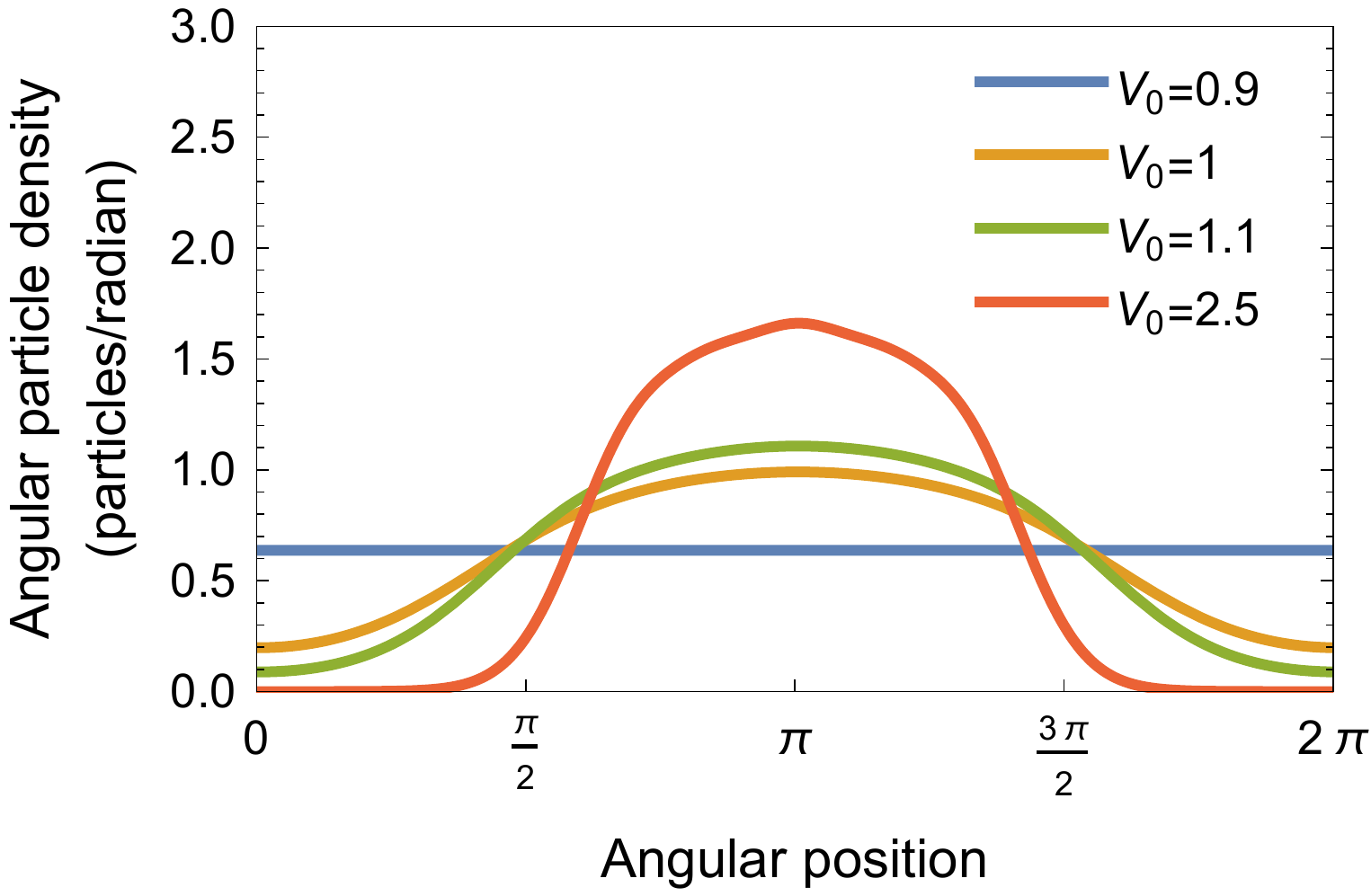}
  \end{center}
\caption{\label{V0varyN2density}Same as Fig.~\ref{V0varyN1density} for the `Nuclear 2' interaction ($R=2$, $N=4$).}
\end{figure}

Finally, Figs.~\ref{V0varyN2} and~\ref{V0varyN2density} show the $V_0$ dependence of the ground-state energy and the angular particle densities for the BSHF method, respectively, for the `Nuclear 2' interaction. Like the previous interactions, the USHF energy has a linear dependence. There is also a clear phase transition from no symmetry breaking to bunching for the other three methods at $V_0\gtrsim1$, as is evident by their energies falling below the USHF energy,  the sharp increase in performance of the BSHF and RSHF methods, and the transition from flat density (indicating no rotational symmetry breaking) to angular densities consisting of a single peak localised to one region of the ring. 

For this interaction, both the BSHF and RSHF methods perform well for all $V_0>1$, with the RSHF only giving a small improvement to the BSHF method. 
This is likely due to the fact that the `Nuclear 2' interaction has a weaker hard-core repulsion
leading to a lack of `spreading out' correlations in the BSHF state. 
As a result, only one phase transition is observed for this case. 

\section{\label{c}Conclusion}

The Hartree-Fock method with broken and restored rotational symmetry has been benchmarked using the ground-state of a quantum ring of identical interacting fermions.
Model interactions were used for simplicity to simulate basic properties of nuclear interactions, such as their hard core repulsion and longer range attraction. 
Therefore, dynamical correlations that are naturally included in EDF methods were neglected.
The focus of this work was on the study of static correlations using symmetry breaking and restoration techniques. 

It was found that rotational symmetry breaking in the Hartree-Fock method brought little improvement to weakly repulsive systems where the spatial correlations between particles were low. As the strength of the interaction increased, so did the strength of correlations and the relative effectiveness of symmetry breaking. This suggests that interaction strength is an important consideration for the use of symmetry breaking in mean-field methods for repulsive systems. For all repulsive systems, symmetry restoration was able to improve upon the symmetry-broken solution. Thus, if high accuracy is required in these systems, implementing a rotational symmetry restoration is likely to be worth doing.

In contrast, it was found that symmetry breaking at the mean-field level was necessary in almost all attractive systems provided the ring was large enough for the particles to form a self-bound system. In general, the restoration of symmetry in these systems provided only small improvements with an exception in the case of the `Nuclear 1' interaction potential which also included a strong hard-core repulsion. 
In this case, the particles become localised (`spreading out') within the self-bound system, and symmetry restoration becomes even more efficient in accounting for the remaining correlation energy. 

Although in the present case of one-dimensional quantum rings, rotation and translation along the ring are equivalent, this is not true for other systems. 
More generally it would be interesting to investigate the effectiveness of the symmetry restoration technique for other transformations using quantum rings or other simplified systems. 
For instance, one could study pairing correlations induced by breaking and restoring U(1) gauge invariance associated with particle number conservation.
These correlations are responsible for nuclear superfluidity and are commonly included through BCS and Hartree-Fock-Bogoliubov approaches.
Here, particle number projection is used to restore the broken symmetry in a beyond mean-field fashion. 

Beyond the inclusion of correlation energy, symmetry restoration through projection techniques presents the advantage of producing states with well-defined quantum numbers (such as the total angular momentum after rotational symmetry restoration).
This is often important for observables requiring a description of the system in the laboratory frame rather than in its intrinsic frame.
Examples include calculation of transition amplitudes in nuclear decay as well as particles scattering off nuclei.

A major caveat to the present study lies in the dimensionality of the system. It has been shown that the correlation energy of a many-electron spherium has a significant dependence on the dimensionality of the system \cite{loosTwoElectronsHypersphere2009}. It would then be natural to assume that dimensionality would also change the effectiveness of the approximate methods studied in this work. Thus, a possible direction for further research is to extend this study to higher dimensional quantum rings, in particular $\mathcal{D}=3$-dimensional glomium which was shown to be the most similar to more realistic models of three-dimensional systems \cite{loosTwoElectronsHypersphere2009} . 

Last but not least, the results presented here only provide a qualitative indication on the static correlations that can be incorporated in nuclear systems. 
The effectiveness of the symmetry breaking and restoration techniques needs to be evaluated with approaches that account for dynamical correlations, i.e., within the EDF approach, or with shell model effective interactions in which the effect of the hard core repulsion has been renormalised. 

\begin{acknowledgments}
This work has been supported by the Australian Research Council through the Discovery Project scheme (project number DP190100256) and through the Centre of Excellence for Dark Matter Particle Physics (CE200100008). Computational resources were provided by the Australian Government through the National Computational Infrastructure (NCI) under the ANU Merit Allocation Scheme and the National Committee Merit Allocation Scheme. J.~Cesca
acknowledges the support of the Australian National University through the Dunbar Physics Honours Scholarship.
\end{acknowledgments}

\bibliography{RingiumPaper}

\begin{thebibliography}{62}%
\makeatletter
\providecommand \@ifxundefined [1]{%
 \@ifx{#1\undefined}
}%
\providecommand \@ifnum [1]{%
 \ifnum #1\expandafter \@firstoftwo
 \else \expandafter \@secondoftwo
 \fi
}%
\providecommand \@ifx [1]{%
 \ifx #1\expandafter \@firstoftwo
 \else \expandafter \@secondoftwo
 \fi
}%
\providecommand \natexlab [1]{#1}%
\providecommand \enquote  [1]{``#1''}%
\providecommand \bibnamefont  [1]{#1}%
\providecommand \bibfnamefont [1]{#1}%
\providecommand \citenamefont [1]{#1}%
\providecommand \href@noop [0]{\@secondoftwo}%
\providecommand \href [0]{\begingroup \@sanitize@url \@href}%
\providecommand \@href[1]{\@@startlink{#1}\@@href}%
\providecommand \@@href[1]{\endgroup#1\@@endlink}%
\providecommand \@sanitize@url [0]{\catcode `\\12\catcode `\$12\catcode
  `\&12\catcode `\#12\catcode `\^12\catcode `\_12\catcode `\%12\relax}%
\providecommand \@@startlink[1]{}%
\providecommand \@@endlink[0]{}%
\providecommand \url  [0]{\begingroup\@sanitize@url \@url }%
\providecommand \@url [1]{\endgroup\@href {#1}{\urlprefix }}%
\providecommand \urlprefix  [0]{URL }%
\providecommand \Eprint [0]{\href }%
\providecommand \doibase [0]{https://doi.org/}%
\providecommand \selectlanguage [0]{\@gobble}%
\providecommand \bibinfo  [0]{\@secondoftwo}%
\providecommand \bibfield  [0]{\@secondoftwo}%
\providecommand \translation [1]{[#1]}%
\providecommand \BibitemOpen [0]{}%
\providecommand \bibitemStop [0]{}%
\providecommand \bibitemNoStop [0]{.\EOS\space}%
\providecommand \EOS [0]{\spacefactor3000\relax}%
\providecommand \BibitemShut  [1]{\csname bibitem#1\endcsname}%
\let\auto@bib@innerbib\@empty
\bibitem [{\citenamefont {Bender}\ \emph {et~al.}(2003)\citenamefont {Bender},
  \citenamefont {Heenen},\ and\ \citenamefont
  {Reinhard}}]{benderSelfconsistentMeanfieldModels2003}%
  \BibitemOpen
  \bibfield  {author} {\bibinfo {author} {\bibfnamefont {M.}~\bibnamefont
  {Bender}}, \bibinfo {author} {\bibfnamefont {P.-H.}\ \bibnamefont {Heenen}},\
  and\ \bibinfo {author} {\bibfnamefont {P.-G.}\ \bibnamefont {Reinhard}},\
  }\bibfield  {title} {\bibinfo {title} {Self-consistent mean-field models for
  nuclear structure},\ }\href {https://doi.org/10.1103/RevModPhys.75.121}
  {\bibfield  {journal} {\bibinfo  {journal} {Rev. Mod. Phys.}\ }\textbf
  {\bibinfo {volume} {75}},\ \bibinfo {pages} {121} (\bibinfo {year}
  {2003})}\BibitemShut {NoStop}%
\bibitem [{\citenamefont {P{\'e}ru}\ and\ \citenamefont
  {Martini}(2014)}]{peru2014}%
  \BibitemOpen
  \bibfield  {author} {\bibinfo {author} {\bibfnamefont {S.}~\bibnamefont
  {P{\'e}ru}}\ and\ \bibinfo {author} {\bibfnamefont {M.}~\bibnamefont
  {Martini}},\ }\bibfield  {title} {\bibinfo {title} {Mean field based
  calculations with the {G}ogny force: Some theoretical tools to explore the
  nuclear structure},\ }\href {https://doi.org/10.1140/epja/i2014-14088-7}
  {\bibfield  {journal} {\bibinfo  {journal} {The European Physical Journal A}\
  }\textbf {\bibinfo {volume} {50}},\ \bibinfo {pages} {88} (\bibinfo {year}
  {2014})}\BibitemShut {NoStop}%
\bibitem [{\citenamefont {Simenel}(2012)}]{simenel2012}%
  \BibitemOpen
  \bibfield  {author} {\bibinfo {author} {\bibfnamefont {C.}~\bibnamefont
  {Simenel}},\ }\bibfield  {title} {\bibinfo {title} {{N}uclear quantum
  many-body dynamics},\ }\href {https://doi.org/10.1140/epja/i2012-12152-0}
  {\bibfield  {journal} {\bibinfo  {journal} {Eur. Phys. J. A}\ }\textbf
  {\bibinfo {volume} {48}},\ \bibinfo {pages} {152} (\bibinfo {year}
  {2012})}\BibitemShut {NoStop}%
\bibitem [{\citenamefont {Simenel}\ and\ \citenamefont
  {Umar}(2018)}]{simenel2018}%
  \BibitemOpen
  \bibfield  {author} {\bibinfo {author} {\bibfnamefont {C.}~\bibnamefont
  {Simenel}}\ and\ \bibinfo {author} {\bibfnamefont {A.~S.}\ \bibnamefont
  {Umar}},\ }\bibfield  {title} {\bibinfo {title} {Heavy-ion collisions and
  fission dynamics with the time--dependent {H}artree-{F}ock theory and its
  extensions},\ }\href {https://doi.org/10.1016/j.ppnp.2018.07.002} {\bibfield
  {journal} {\bibinfo  {journal} {Prog. Part. Nucl. Phys.}\ }\textbf {\bibinfo
  {volume} {103}},\ \bibinfo {pages} {19} (\bibinfo {year} {2018})}\BibitemShut
  {NoStop}%
\bibitem [{\citenamefont {Sheikh}\ \emph {et~al.}(2021)\citenamefont {Sheikh},
  \citenamefont {Dobaczewski}, \citenamefont {Ring}, \citenamefont {Robledo},\
  and\ \citenamefont {Yannouleas}}]{sheikhSymmetryRestorationMeanfield2021}%
  \BibitemOpen
  \bibfield  {author} {\bibinfo {author} {\bibfnamefont {J.~A.}\ \bibnamefont
  {Sheikh}}, \bibinfo {author} {\bibfnamefont {J.}~\bibnamefont {Dobaczewski}},
  \bibinfo {author} {\bibfnamefont {P.}~\bibnamefont {Ring}}, \bibinfo {author}
  {\bibfnamefont {L.~M.}\ \bibnamefont {Robledo}},\ and\ \bibinfo {author}
  {\bibfnamefont {C.}~\bibnamefont {Yannouleas}},\ }\bibfield  {title}
  {\bibinfo {title} {Symmetry restoration in mean-field approaches},\ }\href
  {https://doi.org/10.1088/1361-6471/ac288a} {\bibfield  {journal} {\bibinfo
  {journal} {J. Phys. G: Nucl. Part. Phys.}\ }\textbf {\bibinfo {volume}
  {48}},\ \bibinfo {pages} {123001} (\bibinfo {year} {2021})}\BibitemShut
  {NoStop}%
\bibitem [{\citenamefont {Qiu}\ \emph {et~al.}(2017{\natexlab{a}})\citenamefont
  {Qiu}, \citenamefont {Henderson}, \citenamefont {Zhao},\ and\ \citenamefont
  {Scuseria}}]{qiu2017a}%
  \BibitemOpen
  \bibfield  {author} {\bibinfo {author} {\bibfnamefont {Y.}~\bibnamefont
  {Qiu}}, \bibinfo {author} {\bibfnamefont {T.~M.}\ \bibnamefont {Henderson}},
  \bibinfo {author} {\bibfnamefont {J.}~\bibnamefont {Zhao}},\ and\ \bibinfo
  {author} {\bibfnamefont {G.~E.}\ \bibnamefont {Scuseria}},\ }\bibfield
  {title} {\bibinfo {title} {{Projected coupled cluster theory}},\ }\bibfield
  {journal} {\bibinfo  {journal} {The Journal of Chemical Physics}\ }\textbf
  {\bibinfo {volume} {147}},\ \href {https://doi.org/10.1063/1.4991020}
  {10.1063/1.4991020} (\bibinfo {year} {2017}{\natexlab{a}}),\ \bibinfo {note}
  {064111},\ \Eprint
  {https://arxiv.org/abs/https://pubs.aip.org/aip/jcp/article-pdf/doi/10.1063/1.4991020/16703567/064111\_1\_online.pdf}
  {https://pubs.aip.org/aip/jcp/article-pdf/doi/10.1063/1.4991020/16703567/064111\_1\_online.pdf}
  \BibitemShut {NoStop}%
\bibitem [{\citenamefont {Qiu}\ \emph {et~al.}(2017{\natexlab{b}})\citenamefont
  {Qiu}, \citenamefont {Henderson},\ and\ \citenamefont {Scuseria}}]{qiu2017b}%
  \BibitemOpen
  \bibfield  {author} {\bibinfo {author} {\bibfnamefont {Y.}~\bibnamefont
  {Qiu}}, \bibinfo {author} {\bibfnamefont {T.~M.}\ \bibnamefont {Henderson}},\
  and\ \bibinfo {author} {\bibfnamefont {G.~E.}\ \bibnamefont {Scuseria}},\
  }\bibfield  {title} {\bibinfo {title} {{Projected Hartree-Fock theory as a
  polynomial of particle-hole excitations and its combination with variational
  coupled cluster theory}},\ }\bibfield  {journal} {\bibinfo  {journal} {The
  Journal of Chemical Physics}\ }\textbf {\bibinfo {volume} {146}},\ \href
  {https://doi.org/10.1063/1.4983065} {10.1063/1.4983065} (\bibinfo {year}
  {2017}{\natexlab{b}}),\ \bibinfo {note} {184105},\ \Eprint
  {https://arxiv.org/abs/https://pubs.aip.org/aip/jcp/article-pdf/doi/10.1063/1.4983065/15525823/184105\_1\_online.pdf}
  {https://pubs.aip.org/aip/jcp/article-pdf/doi/10.1063/1.4983065/15525823/184105\_1\_online.pdf}
  \BibitemShut {NoStop}%
\bibitem [{\citenamefont {Duguet}(2014)}]{duguet2015}%
  \BibitemOpen
  \bibfield  {author} {\bibinfo {author} {\bibfnamefont {T.}~\bibnamefont
  {Duguet}},\ }\bibfield  {title} {\bibinfo {title} {Symmetry broken and
  restored coupled-cluster theory: I. {R}otational symmetry and angular
  momentum},\ }\href {https://doi.org/10.1088/0954-3899/42/2/025107} {\bibfield
   {journal} {\bibinfo  {journal} {Journal of Physics G: Nuclear and Particle
  Physics}\ }\textbf {\bibinfo {volume} {42}},\ \bibinfo {pages} {025107}
  (\bibinfo {year} {2014})}\BibitemShut {NoStop}%
\bibitem [{\citenamefont {Duguet}\ and\ \citenamefont
  {Signoracci}(2016)}]{duguet2017}%
  \BibitemOpen
  \bibfield  {author} {\bibinfo {author} {\bibfnamefont {T.}~\bibnamefont
  {Duguet}}\ and\ \bibinfo {author} {\bibfnamefont {A.}~\bibnamefont
  {Signoracci}},\ }\bibfield  {title} {\bibinfo {title} {Symmetry broken and
  restored coupled-cluster theory: I{I}. {G}lobal gauge symmetry and particle
  number},\ }\href {https://doi.org/10.1088/0954-3899/44/1/015103} {\bibfield
  {journal} {\bibinfo  {journal} {Journal of Physics G: Nuclear and Particle
  Physics}\ }\textbf {\bibinfo {volume} {44}},\ \bibinfo {pages} {015103}
  (\bibinfo {year} {2016})}\BibitemShut {NoStop}%
\bibitem [{\citenamefont {Hagen}\ \emph {et~al.}(2022)\citenamefont {Hagen},
  \citenamefont {Novario}, \citenamefont {Sun}, \citenamefont {Papenbrock},
  \citenamefont {Jansen}, \citenamefont {Lietz}, \citenamefont {Duguet},\ and\
  \citenamefont {Tichai}}]{hagen2022}%
  \BibitemOpen
  \bibfield  {author} {\bibinfo {author} {\bibfnamefont {G.}~\bibnamefont
  {Hagen}}, \bibinfo {author} {\bibfnamefont {S.~J.}\ \bibnamefont {Novario}},
  \bibinfo {author} {\bibfnamefont {Z.~H.}\ \bibnamefont {Sun}}, \bibinfo
  {author} {\bibfnamefont {T.}~\bibnamefont {Papenbrock}}, \bibinfo {author}
  {\bibfnamefont {G.~R.}\ \bibnamefont {Jansen}}, \bibinfo {author}
  {\bibfnamefont {J.~G.}\ \bibnamefont {Lietz}}, \bibinfo {author}
  {\bibfnamefont {T.}~\bibnamefont {Duguet}},\ and\ \bibinfo {author}
  {\bibfnamefont {A.}~\bibnamefont {Tichai}},\ }\bibfield  {title} {\bibinfo
  {title} {Angular-momentum projection in coupled-cluster theory: Structure of
  $^{34}\mathrm{Mg}$},\ }\href {https://doi.org/10.1103/PhysRevC.105.064311}
  {\bibfield  {journal} {\bibinfo  {journal} {Phys. Rev. C}\ }\textbf {\bibinfo
  {volume} {105}},\ \bibinfo {pages} {064311} (\bibinfo {year}
  {2022})}\BibitemShut {NoStop}%
\bibitem [{\citenamefont {Lauber}\ \emph {et~al.}(2021)\citenamefont {Lauber},
  \citenamefont {Frye},\ and\ \citenamefont {Johnson}}]{lauber2021}%
  \BibitemOpen
  \bibfield  {author} {\bibinfo {author} {\bibfnamefont {S.~M.}\ \bibnamefont
  {Lauber}}, \bibinfo {author} {\bibfnamefont {H.~C.}\ \bibnamefont {Frye}},\
  and\ \bibinfo {author} {\bibfnamefont {C.~W.}\ \bibnamefont {Johnson}},\
  }\bibfield  {title} {\bibinfo {title} {Benchmarking angular-momentum
  projected {H}artree-{F}ock as an approximation},\ }\href
  {https://doi.org/10.1088/1361-6471/ac1390} {\bibfield  {journal} {\bibinfo
  {journal} {Journal of Physics G: Nuclear and Particle Physics}\ }\textbf
  {\bibinfo {volume} {48}},\ \bibinfo {pages} {095107} (\bibinfo {year}
  {2021})}\BibitemShut {NoStop}%
\bibitem [{\citenamefont {Robledo}\ \emph {et~al.}(2018)\citenamefont
  {Robledo}, \citenamefont {Rodriguez},\ and\ \citenamefont
  {Rodriguez-Guzman}}]{robledo2019}%
  \BibitemOpen
  \bibfield  {author} {\bibinfo {author} {\bibfnamefont {L.~M.}\ \bibnamefont
  {Robledo}}, \bibinfo {author} {\bibfnamefont {T.~R.}\ \bibnamefont
  {Rodriguez}},\ and\ \bibinfo {author} {\bibfnamefont {R.~R.}\ \bibnamefont
  {Rodriguez-Guzman}},\ }\bibfield  {title} {\bibinfo {title} {Mean field and
  beyond description of nuclear structure with the {G}ogny force: a review},\
  }\href {https://doi.org/10.1088/1361-6471/aadebd} {\bibfield  {journal}
  {\bibinfo  {journal} {Journal of Physics G: Nuclear and Particle Physics}\
  }\textbf {\bibinfo {volume} {46}},\ \bibinfo {pages} {013001} (\bibinfo
  {year} {2018})}\BibitemShut {NoStop}%
\bibitem [{\citenamefont {Leonhardt}\ \emph {et~al.}(2020)\citenamefont
  {Leonhardt}, \citenamefont {Pospiech}, \citenamefont {Schallmo},
  \citenamefont {Braun}, \citenamefont {Drischler}, \citenamefont {Hebeler},\
  and\ \citenamefont {Schwenk}}]{leonhardt2020}%
  \BibitemOpen
  \bibfield  {author} {\bibinfo {author} {\bibfnamefont {M.}~\bibnamefont
  {Leonhardt}}, \bibinfo {author} {\bibfnamefont {M.}~\bibnamefont {Pospiech}},
  \bibinfo {author} {\bibfnamefont {B.}~\bibnamefont {Schallmo}}, \bibinfo
  {author} {\bibfnamefont {J.}~\bibnamefont {Braun}}, \bibinfo {author}
  {\bibfnamefont {C.}~\bibnamefont {Drischler}}, \bibinfo {author}
  {\bibfnamefont {K.}~\bibnamefont {Hebeler}},\ and\ \bibinfo {author}
  {\bibfnamefont {A.}~\bibnamefont {Schwenk}},\ }\bibfield  {title} {\bibinfo
  {title} {Symmetric nuclear matter from the strong interaction},\ }\href
  {https://doi.org/10.1103/PhysRevLett.125.142502} {\bibfield  {journal}
  {\bibinfo  {journal} {Phys. Rev. Lett.}\ }\textbf {\bibinfo {volume} {125}},\
  \bibinfo {pages} {142502} (\bibinfo {year} {2020})}\BibitemShut {NoStop}%
\bibitem [{\citenamefont {Atkinson}\ \emph {et~al.}(2020)\citenamefont
  {Atkinson}, \citenamefont {Dickhoff}, \citenamefont {Piarulli}, \citenamefont
  {Rios},\ and\ \citenamefont
  {Wiringa}}]{atkinsonReexaminingRelationBinding2020}%
  \BibitemOpen
  \bibfield  {author} {\bibinfo {author} {\bibfnamefont {M.~C.}\ \bibnamefont
  {Atkinson}}, \bibinfo {author} {\bibfnamefont {W.~H.}\ \bibnamefont
  {Dickhoff}}, \bibinfo {author} {\bibfnamefont {M.}~\bibnamefont {Piarulli}},
  \bibinfo {author} {\bibfnamefont {A.}~\bibnamefont {Rios}},\ and\ \bibinfo
  {author} {\bibfnamefont {R.~B.}\ \bibnamefont {Wiringa}},\ }\bibfield
  {title} {\bibinfo {title} {Reexamining the relation between the binding
  energy of finite nuclei and the equation of state of infinite nuclear
  matter},\ }\href {https://doi.org/10.1103/PhysRevC.102.044333} {\bibfield
  {journal} {\bibinfo  {journal} {Phys. Rev. C}\ }\textbf {\bibinfo {volume}
  {102}},\ \bibinfo {pages} {044333} (\bibinfo {year} {2020})}\BibitemShut
  {NoStop}%
\bibitem [{\citenamefont {Oertel}\ \emph {et~al.}(2017)\citenamefont {Oertel},
  \citenamefont {Hempel}, \citenamefont {Kl{\"a}hn},\ and\ \citenamefont
  {Typel}}]{oertelEquationsStateSupernovae2017}%
  \BibitemOpen
  \bibfield  {author} {\bibinfo {author} {\bibfnamefont {M.}~\bibnamefont
  {Oertel}}, \bibinfo {author} {\bibfnamefont {M.}~\bibnamefont {Hempel}},
  \bibinfo {author} {\bibfnamefont {T.}~\bibnamefont {Kl{\"a}hn}},\ and\
  \bibinfo {author} {\bibfnamefont {S.}~\bibnamefont {Typel}},\ }\bibfield
  {title} {\bibinfo {title} {Equations of state for supernovae and compact
  stars},\ }\href {https://doi.org/10.1103/RevModPhys.89.015007} {\bibfield
  {journal} {\bibinfo  {journal} {Rev. Mod. Phys.}\ }\textbf {\bibinfo {volume}
  {89}},\ \bibinfo {pages} {015007} (\bibinfo {year} {2017})}\BibitemShut
  {NoStop}%
\bibitem [{\citenamefont {Stone}\ and\ \citenamefont
  {Reinhard}(2007)}]{stoneSkyrmeInteractionFinite2007}%
  \BibitemOpen
  \bibfield  {author} {\bibinfo {author} {\bibfnamefont {J.~R.}\ \bibnamefont
  {Stone}}\ and\ \bibinfo {author} {\bibfnamefont {P.-G.}\ \bibnamefont
  {Reinhard}},\ }\bibfield  {title} {\bibinfo {title} {The {{Skyrme
  Interaction}} in finite nuclei and nuclear matter},\ }\href
  {https://doi.org/10.1016/j.ppnp.2006.07.001} {\bibfield  {journal} {\bibinfo
  {journal} {Prog. Part. Nucl. Phys.}\ }\textbf {\bibinfo {volume} {58}},\
  \bibinfo {pages} {587} (\bibinfo {year} {2007})}\BibitemShut {NoStop}%
\bibitem [{\citenamefont {Smerzi}\ \emph {et~al.}(1997)\citenamefont {Smerzi},
  \citenamefont {Ravenhall},\ and\ \citenamefont
  {Pandharipande}}]{smerziNeutronDropsNeutron1997}%
  \BibitemOpen
  \bibfield  {author} {\bibinfo {author} {\bibfnamefont {A.}~\bibnamefont
  {Smerzi}}, \bibinfo {author} {\bibfnamefont {D.~G.}\ \bibnamefont
  {Ravenhall}},\ and\ \bibinfo {author} {\bibfnamefont {V.~R.}\ \bibnamefont
  {Pandharipande}},\ }\bibfield  {title} {\bibinfo {title} {Neutron drops and
  neutron pairing energy},\ }\href {https://doi.org/10.1103/PhysRevC.56.2549}
  {\bibfield  {journal} {\bibinfo  {journal} {Phys. Rev. C}\ }\textbf {\bibinfo
  {volume} {56}},\ \bibinfo {pages} {2549} (\bibinfo {year}
  {1997})}\BibitemShut {NoStop}%
\bibitem [{\citenamefont {Bogner}\ \emph {et~al.}(2011)\citenamefont {Bogner},
  \citenamefont {Furnstahl}, \citenamefont {Hergert}, \citenamefont
  {Kortelainen}, \citenamefont {Maris}, \citenamefont {Stoitsov},\ and\
  \citenamefont {Vary}}]{bogner2011}%
  \BibitemOpen
  \bibfield  {author} {\bibinfo {author} {\bibfnamefont {S.~K.}\ \bibnamefont
  {Bogner}}, \bibinfo {author} {\bibfnamefont {R.~J.}\ \bibnamefont
  {Furnstahl}}, \bibinfo {author} {\bibfnamefont {H.}~\bibnamefont {Hergert}},
  \bibinfo {author} {\bibfnamefont {M.}~\bibnamefont {Kortelainen}}, \bibinfo
  {author} {\bibfnamefont {P.}~\bibnamefont {Maris}}, \bibinfo {author}
  {\bibfnamefont {M.}~\bibnamefont {Stoitsov}},\ and\ \bibinfo {author}
  {\bibfnamefont {J.~P.}\ \bibnamefont {Vary}},\ }\bibfield  {title} {\bibinfo
  {title} {Testing the density matrix expansion against ab initio calculations
  of trapped neutron drops},\ }\href
  {https://doi.org/10.1103/PhysRevC.84.044306} {\bibfield  {journal} {\bibinfo
  {journal} {Phys. Rev. C}\ }\textbf {\bibinfo {volume} {84}},\ \bibinfo
  {pages} {044306} (\bibinfo {year} {2011})}\BibitemShut {NoStop}%
\bibitem [{\citenamefont {Potter}\ \emph {et~al.}(2014)\citenamefont {Potter},
  \citenamefont {Fischer}, \citenamefont {Maris}, \citenamefont {Vary},
  \citenamefont {Binder}, \citenamefont {Calci}, \citenamefont {Langhammer},\
  and\ \citenamefont {Roth}}]{potter2014}%
  \BibitemOpen
  \bibfield  {author} {\bibinfo {author} {\bibfnamefont {H.}~\bibnamefont
  {Potter}}, \bibinfo {author} {\bibfnamefont {S.}~\bibnamefont {Fischer}},
  \bibinfo {author} {\bibfnamefont {P.}~\bibnamefont {Maris}}, \bibinfo
  {author} {\bibfnamefont {J.}~\bibnamefont {Vary}}, \bibinfo {author}
  {\bibfnamefont {S.}~\bibnamefont {Binder}}, \bibinfo {author} {\bibfnamefont
  {A.}~\bibnamefont {Calci}}, \bibinfo {author} {\bibfnamefont
  {J.}~\bibnamefont {Langhammer}},\ and\ \bibinfo {author} {\bibfnamefont
  {R.}~\bibnamefont {Roth}},\ }\bibfield  {title} {\bibinfo {title} {Ab initio
  study of neutron drops with chiral {H}amiltonians},\ }\href
  {https://doi.org/https://doi.org/10.1016/j.physletb.2014.10.020} {\bibfield
  {journal} {\bibinfo  {journal} {Physics Letters B}\ }\textbf {\bibinfo
  {volume} {739}},\ \bibinfo {pages} {445 } (\bibinfo {year}
  {2014})}\BibitemShut {NoStop}%
\bibitem [{\citenamefont {Shen}\ \emph {et~al.}(2018)\citenamefont {Shen},
  \citenamefont {Liang}, \citenamefont {Meng}, \citenamefont {Ring},\ and\
  \citenamefont {Zhang}}]{shen2018}%
  \BibitemOpen
  \bibfield  {author} {\bibinfo {author} {\bibfnamefont {S.}~\bibnamefont
  {Shen}}, \bibinfo {author} {\bibfnamefont {H.}~\bibnamefont {Liang}},
  \bibinfo {author} {\bibfnamefont {J.}~\bibnamefont {Meng}}, \bibinfo {author}
  {\bibfnamefont {P.}~\bibnamefont {Ring}},\ and\ \bibinfo {author}
  {\bibfnamefont {S.}~\bibnamefont {Zhang}},\ }\bibfield  {title} {\bibinfo
  {title} {Relativistic {B}rueckner-{H}artree-{F}ock theory for neutron
  drops},\ }\href {https://doi.org/10.1103/PhysRevC.97.054312} {\bibfield
  {journal} {\bibinfo  {journal} {Phys. Rev. C}\ }\textbf {\bibinfo {volume}
  {97}},\ \bibinfo {pages} {054312} (\bibinfo {year} {2018})}\BibitemShut
  {NoStop}%
\bibitem [{\citenamefont {Zhao}\ \emph {et~al.}(2020)\citenamefont {Zhao},
  \citenamefont {Zhao},\ and\ \citenamefont {Meng}}]{zhao2020}%
  \BibitemOpen
  \bibfield  {author} {\bibinfo {author} {\bibfnamefont {Q.}~\bibnamefont
  {Zhao}}, \bibinfo {author} {\bibfnamefont {P.}~\bibnamefont {Zhao}},\ and\
  \bibinfo {author} {\bibfnamefont {J.}~\bibnamefont {Meng}},\ }\bibfield
  {title} {\bibinfo {title} {Impact of tensor forces on spin-orbit splittings
  in neutron-proton drops},\ }\href
  {https://doi.org/10.1103/PhysRevC.102.034322} {\bibfield  {journal} {\bibinfo
   {journal} {Phys. Rev. C}\ }\textbf {\bibinfo {volume} {102}},\ \bibinfo
  {pages} {034322} (\bibinfo {year} {2020})}\BibitemShut {NoStop}%
\bibitem [{\citenamefont {Pudliner}\ \emph {et~al.}(1996)\citenamefont
  {Pudliner}, \citenamefont {Smerzi}, \citenamefont {Carlson}, \citenamefont
  {Pandharipande}, \citenamefont {Pieper},\ and\ \citenamefont
  {Ravenhall}}]{pudlinerNeutronDropsSkyrme1996}%
  \BibitemOpen
  \bibfield  {author} {\bibinfo {author} {\bibfnamefont {B.~S.}\ \bibnamefont
  {Pudliner}}, \bibinfo {author} {\bibfnamefont {A.}~\bibnamefont {Smerzi}},
  \bibinfo {author} {\bibfnamefont {J.}~\bibnamefont {Carlson}}, \bibinfo
  {author} {\bibfnamefont {V.~R.}\ \bibnamefont {Pandharipande}}, \bibinfo
  {author} {\bibfnamefont {S.~C.}\ \bibnamefont {Pieper}},\ and\ \bibinfo
  {author} {\bibfnamefont {D.~G.}\ \bibnamefont {Ravenhall}},\ }\bibfield
  {title} {\bibinfo {title} {Neutron {{Drops}} and {{Skyrme Energy-Density
  Functionals}}},\ }\href {https://doi.org/10.1103/PhysRevLett.76.2416}
  {\bibfield  {journal} {\bibinfo  {journal} {Phys. Rev. Lett.}\ }\textbf
  {\bibinfo {volume} {76}},\ \bibinfo {pages} {2416} (\bibinfo {year}
  {1996})}\BibitemShut {NoStop}%
\bibitem [{\citenamefont {Gandolfi}\ \emph {et~al.}(2011)\citenamefont
  {Gandolfi}, \citenamefont {Carlson},\ and\ \citenamefont
  {Pieper}}]{gandolfi2011}%
  \BibitemOpen
  \bibfield  {author} {\bibinfo {author} {\bibfnamefont {S.}~\bibnamefont
  {Gandolfi}}, \bibinfo {author} {\bibfnamefont {J.}~\bibnamefont {Carlson}},\
  and\ \bibinfo {author} {\bibfnamefont {S.~C.}\ \bibnamefont {Pieper}},\
  }\bibfield  {title} {\bibinfo {title} {Cold neutrons trapped in external
  fields},\ }\href {https://doi.org/10.1103/PhysRevLett.106.012501} {\bibfield
  {journal} {\bibinfo  {journal} {Phys. Rev. Lett.}\ }\textbf {\bibinfo
  {volume} {106}},\ \bibinfo {pages} {012501} (\bibinfo {year}
  {2011})}\BibitemShut {NoStop}%
\bibitem [{\citenamefont {Maris}\ \emph {et~al.}(2013)\citenamefont {Maris},
  \citenamefont {Vary}, \citenamefont {Gandolfi}, \citenamefont {Carlson},\
  and\ \citenamefont {Pieper}}]{maris2013}%
  \BibitemOpen
  \bibfield  {author} {\bibinfo {author} {\bibfnamefont {P.}~\bibnamefont
  {Maris}}, \bibinfo {author} {\bibfnamefont {J.~P.}\ \bibnamefont {Vary}},
  \bibinfo {author} {\bibfnamefont {S.}~\bibnamefont {Gandolfi}}, \bibinfo
  {author} {\bibfnamefont {J.}~\bibnamefont {Carlson}},\ and\ \bibinfo {author}
  {\bibfnamefont {S.~C.}\ \bibnamefont {Pieper}},\ }\bibfield  {title}
  {\bibinfo {title} {Properties of trapped neutrons interacting with realistic
  nuclear {H}amiltonians},\ }\href {https://doi.org/10.1103/PhysRevC.87.054318}
  {\bibfield  {journal} {\bibinfo  {journal} {Phys. Rev. C}\ }\textbf {\bibinfo
  {volume} {87}},\ \bibinfo {pages} {054318} (\bibinfo {year}
  {2013})}\BibitemShut {NoStop}%
\bibitem [{\citenamefont {Bonche}\ \emph {et~al.}(1976)\citenamefont {Bonche},
  \citenamefont {Koonin},\ and\ \citenamefont {Negele}}]{bonche1976}%
  \BibitemOpen
  \bibfield  {author} {\bibinfo {author} {\bibfnamefont {P.}~\bibnamefont
  {Bonche}}, \bibinfo {author} {\bibfnamefont {S.}~\bibnamefont {Koonin}},\
  and\ \bibinfo {author} {\bibfnamefont {J.~W.}\ \bibnamefont {Negele}},\
  }\bibfield  {title} {\bibinfo {title} {{O}ne-dimensional nuclear dynamics in
  time-dependent {H}artree-{F}ock approximation},\ }\href
  {https://doi.org/10.1103/PhysRevC.13.1226} {\bibfield  {journal} {\bibinfo
  {journal} {Phys. Rev. C}\ }\textbf {\bibinfo {volume} {13}},\ \bibinfo
  {pages} {1226} (\bibinfo {year} {1976})}\BibitemShut {NoStop}%
\bibitem [{\citenamefont {Rios}\ \emph {et~al.}(2011)\citenamefont {Rios},
  \citenamefont {Barker}, \citenamefont {Buchler},\ and\ \citenamefont
  {Danielewicz}}]{rios2011}%
  \BibitemOpen
  \bibfield  {author} {\bibinfo {author} {\bibfnamefont {A.}~\bibnamefont
  {Rios}}, \bibinfo {author} {\bibfnamefont {B.}~\bibnamefont {Barker}},
  \bibinfo {author} {\bibfnamefont {M.}~\bibnamefont {Buchler}},\ and\ \bibinfo
  {author} {\bibfnamefont {P.}~\bibnamefont {Danielewicz}},\ }\bibfield
  {title} {\bibinfo {title} {Towards a nonequilibrium {G}reen's function
  description of nuclear reactions: One-dimensional mean-field dynamics},\
  }\href {https://doi.org/https://doi.org/10.1016/j.aop.2010.12.009} {\bibfield
   {journal} {\bibinfo  {journal} {Annals of Physics}\ }\textbf {\bibinfo
  {volume} {326}},\ \bibinfo {pages} {1274 } (\bibinfo {year}
  {2011})}\BibitemShut {NoStop}%
\bibitem [{\citenamefont {Simenel}\ and\ \citenamefont
  {Umar}(2014)}]{simenel2014a}%
  \BibitemOpen
  \bibfield  {author} {\bibinfo {author} {\bibfnamefont {C.}~\bibnamefont
  {Simenel}}\ and\ \bibinfo {author} {\bibfnamefont {A.~S.}\ \bibnamefont
  {Umar}},\ }\bibfield  {title} {\bibinfo {title} {{F}ormation and dynamics of
  fission fragments},\ }\href {https://doi.org/10.1103/PhysRevC.89.031601}
  {\bibfield  {journal} {\bibinfo  {journal} {Phys. Rev. C}\ }\textbf {\bibinfo
  {volume} {89}},\ \bibinfo {pages} {031601(R)} (\bibinfo {year}
  {2014})}\BibitemShut {NoStop}%
\bibitem [{\citenamefont {Severyukhin}\ \emph {et~al.}(2006)\citenamefont
  {Severyukhin}, \citenamefont {Bender},\ and\ \citenamefont
  {Heenen}}]{severyukhin2006}%
  \BibitemOpen
  \bibfield  {author} {\bibinfo {author} {\bibfnamefont {A.~P.}\ \bibnamefont
  {Severyukhin}}, \bibinfo {author} {\bibfnamefont {M.}~\bibnamefont
  {Bender}},\ and\ \bibinfo {author} {\bibfnamefont {P.-H.}\ \bibnamefont
  {Heenen}},\ }\bibfield  {title} {\bibinfo {title} {Beyond mean field study of
  excited states: Analysis within the {L}ipkin model},\ }\href
  {https://doi.org/10.1103/PhysRevC.74.024311} {\bibfield  {journal} {\bibinfo
  {journal} {Phys. Rev. C}\ }\textbf {\bibinfo {volume} {74}},\ \bibinfo
  {pages} {024311} (\bibinfo {year} {2006})}\BibitemShut {NoStop}%
\bibitem [{\citenamefont {Viefers}\ \emph {et~al.}(2004)\citenamefont
  {Viefers}, \citenamefont {Koskinen}, \citenamefont {Deo},\ and\ \citenamefont
  {Manninen}}]{viefersQuantumRingsBeginners2004}%
  \BibitemOpen
  \bibfield  {author} {\bibinfo {author} {\bibfnamefont {S.}~\bibnamefont
  {Viefers}}, \bibinfo {author} {\bibfnamefont {P.}~\bibnamefont {Koskinen}},
  \bibinfo {author} {\bibfnamefont {P.~S.}\ \bibnamefont {Deo}},\ and\ \bibinfo
  {author} {\bibfnamefont {M.}~\bibnamefont {Manninen}},\ }\bibfield  {title}
  {\bibinfo {title} {Quantum rings for beginners: {{Energy}} spectra and
  persistent currents},\ }\href {https://doi.org/10.1016/j.physe.2003.08.076}
  {\bibfield  {journal} {\bibinfo  {journal} {Physica E Low Dimens. Syst.
  Nanostruct.}\ }\textbf {\bibinfo {volume} {21}},\ \bibinfo {pages} {1}
  (\bibinfo {year} {2004})}\BibitemShut {NoStop}%
\bibitem [{\citenamefont {Loos}\ and\ \citenamefont
  {Gill}(2009)}]{loosTwoElectronsHypersphere2009}%
  \BibitemOpen
  \bibfield  {author} {\bibinfo {author} {\bibfnamefont {P.-F.}\ \bibnamefont
  {Loos}}\ and\ \bibinfo {author} {\bibfnamefont {P.~M.~W.}\ \bibnamefont
  {Gill}},\ }\bibfield  {title} {\bibinfo {title} {Two {{Electrons}} on a
  {{Hypersphere}}: {{A Quasiexactly Solvable Model}}},\ }\href
  {https://doi.org/10.1103/PhysRevLett.103.123008} {\bibfield  {journal}
  {\bibinfo  {journal} {Phys. Rev. Lett.}\ }\textbf {\bibinfo {volume} {103}},\
  \bibinfo {pages} {123008} (\bibinfo {year} {2009})}\BibitemShut {NoStop}%
\bibitem [{\citenamefont {Emperador}\ \emph {et~al.}(2001)\citenamefont
  {Emperador}, \citenamefont {Pi}, \citenamefont {Barranco},\ and\
  \citenamefont {Lipparini}}]{emperador2001}%
  \BibitemOpen
  \bibfield  {author} {\bibinfo {author} {\bibfnamefont {A.}~\bibnamefont
  {Emperador}}, \bibinfo {author} {\bibfnamefont {M.}~\bibnamefont {Pi}},
  \bibinfo {author} {\bibfnamefont {M.}~\bibnamefont {Barranco}},\ and\
  \bibinfo {author} {\bibfnamefont {E.}~\bibnamefont {Lipparini}},\ }\bibfield
  {title} {\bibinfo {title} {Multipole modes and spin features in the {R}aman
  spectrum of nanoscopic quantum rings},\ }\href
  {https://doi.org/10.1103/PhysRevB.64.155304} {\bibfield  {journal} {\bibinfo
  {journal} {Phys. Rev. B}\ }\textbf {\bibinfo {volume} {64}},\ \bibinfo
  {pages} {155304} (\bibinfo {year} {2001})}\BibitemShut {NoStop}%
\bibitem [{\citenamefont {Emperador}\ \emph {et~al.}(2003)\citenamefont
  {Emperador}, \citenamefont {Pederiva},\ and\ \citenamefont
  {Lipparini}}]{emperador2003}%
  \BibitemOpen
  \bibfield  {author} {\bibinfo {author} {\bibfnamefont {A.}~\bibnamefont
  {Emperador}}, \bibinfo {author} {\bibfnamefont {F.}~\bibnamefont
  {Pederiva}},\ and\ \bibinfo {author} {\bibfnamefont {E.}~\bibnamefont
  {Lipparini}},\ }\bibfield  {title} {\bibinfo {title} {Spin- and
  localization-induced fractional {A}haronov-{B}ohm effect},\ }\href
  {https://doi.org/10.1103/PhysRevB.68.115312} {\bibfield  {journal} {\bibinfo
  {journal} {Phys. Rev. B}\ }\textbf {\bibinfo {volume} {68}},\ \bibinfo
  {pages} {115312} (\bibinfo {year} {2003})}\BibitemShut {NoStop}%
\bibitem [{\citenamefont {Zhu}\ \emph {et~al.}(2003)\citenamefont {Zhu},
  \citenamefont {Dai},\ and\ \citenamefont {Hu}}]{zhu2003}%
  \BibitemOpen
  \bibfield  {author} {\bibinfo {author} {\bibfnamefont {J.-L.}\ \bibnamefont
  {Zhu}}, \bibinfo {author} {\bibfnamefont {Z.}~\bibnamefont {Dai}},\ and\
  \bibinfo {author} {\bibfnamefont {X.}~\bibnamefont {Hu}},\ }\bibfield
  {title} {\bibinfo {title} {Two electrons in one-dimensional nanorings: Exact
  solutions and interaction energies},\ }\href
  {https://doi.org/10.1103/PhysRevB.68.045324} {\bibfield  {journal} {\bibinfo
  {journal} {Phys. Rev. B}\ }\textbf {\bibinfo {volume} {68}},\ \bibinfo
  {pages} {045324} (\bibinfo {year} {2003})}\BibitemShut {NoStop}%
\bibitem [{\citenamefont {Fogler}\ and\ \citenamefont
  {Pivovarov}(2005)}]{fogler2005b}%
  \BibitemOpen
  \bibfield  {author} {\bibinfo {author} {\bibfnamefont {M.~M.}\ \bibnamefont
  {Fogler}}\ and\ \bibinfo {author} {\bibfnamefont {E.}~\bibnamefont
  {Pivovarov}},\ }\bibfield  {title} {\bibinfo {title} {Exchange interaction in
  quantum rings and wires in the {W}igner-crystal limit},\ }\href
  {https://doi.org/10.1103/PhysRevB.72.195344} {\bibfield  {journal} {\bibinfo
  {journal} {Phys. Rev. B}\ }\textbf {\bibinfo {volume} {72}},\ \bibinfo
  {pages} {195344} (\bibinfo {year} {2005})}\BibitemShut {NoStop}%
\bibitem [{\citenamefont {Aichinger}\ \emph {et~al.}(2006)\citenamefont
  {Aichinger}, \citenamefont {Chin}, \citenamefont {Krotscheck},\ and\
  \citenamefont {R\"as\"anen}}]{aichinger2006}%
  \BibitemOpen
  \bibfield  {author} {\bibinfo {author} {\bibfnamefont {M.}~\bibnamefont
  {Aichinger}}, \bibinfo {author} {\bibfnamefont {S.~A.}\ \bibnamefont {Chin}},
  \bibinfo {author} {\bibfnamefont {E.}~\bibnamefont {Krotscheck}},\ and\
  \bibinfo {author} {\bibfnamefont {E.}~\bibnamefont {R\"as\"anen}},\
  }\bibfield  {title} {\bibinfo {title} {Effects of geometry and impurities on
  quantum rings in magnetic fields},\ }\href
  {https://doi.org/10.1103/PhysRevB.73.195310} {\bibfield  {journal} {\bibinfo
  {journal} {Phys. Rev. B}\ }\textbf {\bibinfo {volume} {73}},\ \bibinfo
  {pages} {195310} (\bibinfo {year} {2006})}\BibitemShut {NoStop}%
\bibitem [{\citenamefont {Gylfadottir}\ \emph {et~al.}(2006)\citenamefont
  {Gylfadottir}, \citenamefont {Harju}, \citenamefont {Jouttenus},\ and\
  \citenamefont {Webb}}]{gylfadottir2006}%
  \BibitemOpen
  \bibfield  {author} {\bibinfo {author} {\bibfnamefont {S.~S.}\ \bibnamefont
  {Gylfadottir}}, \bibinfo {author} {\bibfnamefont {A.}~\bibnamefont {Harju}},
  \bibinfo {author} {\bibfnamefont {T.}~\bibnamefont {Jouttenus}},\ and\
  \bibinfo {author} {\bibfnamefont {C.}~\bibnamefont {Webb}},\ }\bibfield
  {title} {\bibinfo {title} {Interacting electrons on a quantum ring: exact and
  variational approach},\ }\href {https://doi.org/10.1088/1367-2630/8/9/211}
  {\bibfield  {journal} {\bibinfo  {journal} {New Journal of Physics}\ }\textbf
  {\bibinfo {volume} {8}},\ \bibinfo {pages} {211} (\bibinfo {year}
  {2006})}\BibitemShut {NoStop}%
\bibitem [{\citenamefont {R\"as\"anen}\ \emph {et~al.}(2009)\citenamefont
  {R\"as\"anen}, \citenamefont {Pittalis}, \citenamefont {Proetto},\ and\
  \citenamefont {Gross}}]{rasanen2009}%
  \BibitemOpen
  \bibfield  {author} {\bibinfo {author} {\bibfnamefont {E.}~\bibnamefont
  {R\"as\"anen}}, \bibinfo {author} {\bibfnamefont {S.}~\bibnamefont
  {Pittalis}}, \bibinfo {author} {\bibfnamefont {C.~R.}\ \bibnamefont
  {Proetto}},\ and\ \bibinfo {author} {\bibfnamefont {E.~K.~U.}\ \bibnamefont
  {Gross}},\ }\bibfield  {title} {\bibinfo {title} {Electronic exchange in
  quantum rings: Beyond the local-density approximation},\ }\href
  {https://doi.org/10.1103/PhysRevB.79.121305} {\bibfield  {journal} {\bibinfo
  {journal} {Phys. Rev. B}\ }\textbf {\bibinfo {volume} {79}},\ \bibinfo
  {pages} {121305(R)} (\bibinfo {year} {2009})}\BibitemShut {NoStop}%
\bibitem [{\citenamefont {Manninen}\ and\ \citenamefont
  {Reimann}(2009)}]{manninen2009}%
  \BibitemOpen
  \bibfield  {author} {\bibinfo {author} {\bibfnamefont {M.}~\bibnamefont
  {Manninen}}\ and\ \bibinfo {author} {\bibfnamefont {S.~M.}\ \bibnamefont
  {Reimann}},\ }\bibfield  {title} {\bibinfo {title} {Electron correlation in
  metal clusters, quantum dots and quantum rings},\ }\href
  {https://doi.org/10.1088/1751-8113/42/21/214019} {\bibfield  {journal}
  {\bibinfo  {journal} {Journal of Physics A: Mathematical and Theoretical}\
  }\textbf {\bibinfo {volume} {42}},\ \bibinfo {pages} {214019} (\bibinfo
  {year} {2009})}\BibitemShut {NoStop}%
\bibitem [{\citenamefont {Loos}\ and\ \citenamefont
  {Gill}(2012)}]{loosExactWaveFunctions2012}%
  \BibitemOpen
  \bibfield  {author} {\bibinfo {author} {\bibfnamefont {P.-F.}\ \bibnamefont
  {Loos}}\ and\ \bibinfo {author} {\bibfnamefont {P.~M.~W.}\ \bibnamefont
  {Gill}},\ }\bibfield  {title} {\bibinfo {title} {Exact {{Wave Functions}} of
  {{Two-Electron Quantum Rings}}},\ }\href
  {https://doi.org/10.1103/PhysRevLett.108.083002} {\bibfield  {journal}
  {\bibinfo  {journal} {Phys. Rev. Lett.}\ }\textbf {\bibinfo {volume} {108}},\
  \bibinfo {pages} {083002} (\bibinfo {year} {2012})}\BibitemShut {NoStop}%
\bibitem [{\citenamefont {Loos}\ and\ \citenamefont {Gill}(2013)}]{loos2013}%
  \BibitemOpen
  \bibfield  {author} {\bibinfo {author} {\bibfnamefont {P.-F.}\ \bibnamefont
  {Loos}}\ and\ \bibinfo {author} {\bibfnamefont {P.~M.~W.}\ \bibnamefont
  {Gill}},\ }\bibfield  {title} {\bibinfo {title} {Uniform electron gases. i.
  electrons on a ring},\ }\href {https://doi.org/10.1063/1.4802589} {\bibfield
  {journal} {\bibinfo  {journal} {J. Chem. Phys}\ }\textbf {\bibinfo {volume}
  {138}},\ \bibinfo {pages} {164124} (\bibinfo {year} {2013})}\BibitemShut
  {NoStop}%
\bibitem [{\citenamefont {Rogers}\ and\ \citenamefont
  {Loos}(2017)}]{rogers2017}%
  \BibitemOpen
  \bibfield  {author} {\bibinfo {author} {\bibfnamefont {F.~J.~M.}\
  \bibnamefont {Rogers}}\ and\ \bibinfo {author} {\bibfnamefont {P.-F.}\
  \bibnamefont {Loos}},\ }\bibfield  {title} {\bibinfo {title} {Excited-state
  {W}igner crystals},\ }\href {https://doi.org/10.1063/1.4974839} {\bibfield
  {journal} {\bibinfo  {journal} {The Journal of Chemical Physics}\ }\textbf
  {\bibinfo {volume} {146}},\ \bibinfo {pages} {044114} (\bibinfo {year}
  {2017})},\ \Eprint {https://arxiv.org/abs/https://doi.org/10.1063/1.4974839}
  {https://doi.org/10.1063/1.4974839} \BibitemShut {NoStop}%
\bibitem [{\citenamefont {Li}\ \emph {et~al.}(2021)\citenamefont {Li},
  \citenamefont {Liu},\ and\ \citenamefont {Zhang}}]{li2021}%
  \BibitemOpen
  \bibfield  {author} {\bibinfo {author} {\bibfnamefont {Q.}~\bibnamefont
  {Li}}, \bibinfo {author} {\bibfnamefont {J.-J.}\ \bibnamefont {Liu}},\ and\
  \bibinfo {author} {\bibfnamefont {Y.-T.}\ \bibnamefont {Zhang}},\ }\bibfield
  {title} {\bibinfo {title} {Non-{H}ermitian {A}haronov-{B}ohm effect in the
  quantum ring},\ }\href {https://doi.org/10.1103/PhysRevB.103.035415}
  {\bibfield  {journal} {\bibinfo  {journal} {Phys. Rev. B}\ }\textbf {\bibinfo
  {volume} {103}},\ \bibinfo {pages} {035415} (\bibinfo {year}
  {2021})}\BibitemShut {NoStop}%
\bibitem [{\citenamefont {Neill}\ \emph {et~al.}(2021)\citenamefont {Neill}
  \emph {et~al.}}]{neillAccuratelyComputingElectronic2021a}%
  \BibitemOpen
  \bibfield  {author} {\bibinfo {author} {\bibfnamefont {C.}~\bibnamefont
  {Neill}} \emph {et~al.},\ }\bibfield  {title} {\bibinfo {title} {Accurately
  computing the electronic properties of a quantum ring},\ }\href
  {https://doi.org/10.1038/s41586-021-03576-2} {\bibfield  {journal} {\bibinfo
  {journal} {Nature}\ }\textbf {\bibinfo {volume} {594}},\ \bibinfo {pages}
  {508} (\bibinfo {year} {2021})}\BibitemShut {NoStop}%
\bibitem [{\citenamefont {Hern{\'a}ndez}\ \emph {et~al.}(2022)\citenamefont
  {Hern{\'a}ndez}, \citenamefont {{L{\'o}pez-Doria}}, \citenamefont {Rivera},\
  and\ \citenamefont {Fulla}}]{hernandezRefractiveIndexChange2022}%
  \BibitemOpen
  \bibfield  {author} {\bibinfo {author} {\bibfnamefont {N.}~\bibnamefont
  {Hern{\'a}ndez}}, \bibinfo {author} {\bibfnamefont {R.~A.}\ \bibnamefont
  {{L{\'o}pez-Doria}}}, \bibinfo {author} {\bibfnamefont {I.~E.}\ \bibnamefont
  {Rivera}},\ and\ \bibinfo {author} {\bibfnamefont {M.~R.}\ \bibnamefont
  {Fulla}},\ }\bibfield  {title} {\bibinfo {title} {Refractive index change of
  a {{D2}}+complex in {{GaAs}}/{{AlxGa1}}-{{xAs}} quantum ring},\ }\href
  {https://doi.org/10.1007/s10853-021-06763-8} {\bibfield  {journal} {\bibinfo
  {journal} {J. Mater. Sci.}\ }\textbf {\bibinfo {volume} {57}},\ \bibinfo
  {pages} {8417} (\bibinfo {year} {2022})}\BibitemShut {NoStop}%
\bibitem [{\citenamefont {Judd}\ \emph {et~al.}(2020)\citenamefont {Judd},
  \citenamefont {Nizovtsev}, \citenamefont {Plougmann}, \citenamefont
  {Kondratuk}, \citenamefont {Anderson}, \citenamefont {Besley},\ and\
  \citenamefont {Saywell}}]{juddMolecularQuantumRings2020}%
  \BibitemOpen
  \bibfield  {author} {\bibinfo {author} {\bibfnamefont {C.~J.}\ \bibnamefont
  {Judd}}, \bibinfo {author} {\bibfnamefont {A.~S.}\ \bibnamefont {Nizovtsev}},
  \bibinfo {author} {\bibfnamefont {R.}~\bibnamefont {Plougmann}}, \bibinfo
  {author} {\bibfnamefont {D.~V.}\ \bibnamefont {Kondratuk}}, \bibinfo {author}
  {\bibfnamefont {H.~L.}\ \bibnamefont {Anderson}}, \bibinfo {author}
  {\bibfnamefont {E.}~\bibnamefont {Besley}},\ and\ \bibinfo {author}
  {\bibfnamefont {A.}~\bibnamefont {Saywell}},\ }\bibfield  {title} {\bibinfo
  {title} {Molecular {Quantum} {Rings} {Formed} from a $\pi$-{Conjugated}
  {Macrocycle}},\ }\href {https://doi.org/10.1103/PhysRevLett.125.206803}
  {\bibfield  {journal} {\bibinfo  {journal} {Phys. Rev. Lett.}\ }\textbf
  {\bibinfo {volume} {125}},\ \bibinfo {pages} {206803} (\bibinfo {year}
  {2020})}\BibitemShut {NoStop}%
\bibitem [{\citenamefont {Steck}\ \emph {et~al.}(1996)\citenamefont {Steck},
  \citenamefont {Beckert}, \citenamefont {Eickhoff}, \citenamefont {Franzke},
  \citenamefont {Nolden}, \citenamefont {Reich}, \citenamefont {Schlitt},\ and\
  \citenamefont {Winkler}}]{steck1996}%
  \BibitemOpen
  \bibfield  {author} {\bibinfo {author} {\bibfnamefont {M.}~\bibnamefont
  {Steck}}, \bibinfo {author} {\bibfnamefont {K.}~\bibnamefont {Beckert}},
  \bibinfo {author} {\bibfnamefont {H.}~\bibnamefont {Eickhoff}}, \bibinfo
  {author} {\bibfnamefont {B.}~\bibnamefont {Franzke}}, \bibinfo {author}
  {\bibfnamefont {F.}~\bibnamefont {Nolden}}, \bibinfo {author} {\bibfnamefont
  {H.}~\bibnamefont {Reich}}, \bibinfo {author} {\bibfnamefont
  {B.}~\bibnamefont {Schlitt}},\ and\ \bibinfo {author} {\bibfnamefont
  {T.}~\bibnamefont {Winkler}},\ }\bibfield  {title} {\bibinfo {title}
  {Anomalous temperature reduction of electron-cooled heavy ion beams in the
  storage ring {ESR}},\ }\href {https://doi.org/10.1103/PhysRevLett.77.3803}
  {\bibfield  {journal} {\bibinfo  {journal} {Phys. Rev. Lett.}\ }\textbf
  {\bibinfo {volume} {77}},\ \bibinfo {pages} {3803} (\bibinfo {year}
  {1996})}\BibitemShut {NoStop}%
\bibitem [{\citenamefont {Danared}\ \emph {et~al.}(2002)\citenamefont
  {Danared}, \citenamefont {K\"allberg}, \citenamefont {Rensfelt},\ and\
  \citenamefont {Simonsson}}]{danared2002}%
  \BibitemOpen
  \bibfield  {author} {\bibinfo {author} {\bibfnamefont {H.}~\bibnamefont
  {Danared}}, \bibinfo {author} {\bibfnamefont {A.}~\bibnamefont {K\"allberg}},
  \bibinfo {author} {\bibfnamefont {K.-G.}\ \bibnamefont {Rensfelt}},\ and\
  \bibinfo {author} {\bibfnamefont {A.}~\bibnamefont {Simonsson}},\ }\bibfield
  {title} {\bibinfo {title} {Model and observations of schottky-noise
  suppression in a cold heavy-ion beam},\ }\href
  {https://doi.org/10.1103/PhysRevLett.88.174801} {\bibfield  {journal}
  {\bibinfo  {journal} {Phys. Rev. Lett.}\ }\textbf {\bibinfo {volume} {88}},\
  \bibinfo {pages} {174801} (\bibinfo {year} {2002})}\BibitemShut {NoStop}%
\bibitem [{\citenamefont {Hasse}(2003)}]{hasseStaticCriteriaExistence2003}%
  \BibitemOpen
  \bibfield  {author} {\bibinfo {author} {\bibfnamefont {R.~W.}\ \bibnamefont
  {Hasse}},\ }\bibfield  {title} {\bibinfo {title} {Static criteria for the
  existence of {{Coulomb}} strings in storage rings},\ }\href
  {https://doi.org/10.1103/PhysRevLett.90.204801} {\bibfield  {journal}
  {\bibinfo  {journal} {Phys. Rev. Lett.}\ }\textbf {\bibinfo {volume} {90}},\
  \bibinfo {pages} {204801} (\bibinfo {year} {2003})}\BibitemShut {NoStop}%
\bibitem [{\citenamefont {Bargi}\ \emph {et~al.}(2010)\citenamefont {Bargi},
  \citenamefont {Malet}, \citenamefont {Kavoulakis},\ and\ \citenamefont
  {Reimann}}]{bargi2010}%
  \BibitemOpen
  \bibfield  {author} {\bibinfo {author} {\bibfnamefont {S.}~\bibnamefont
  {Bargi}}, \bibinfo {author} {\bibfnamefont {F.}~\bibnamefont {Malet}},
  \bibinfo {author} {\bibfnamefont {G.~M.}\ \bibnamefont {Kavoulakis}},\ and\
  \bibinfo {author} {\bibfnamefont {S.~M.}\ \bibnamefont {Reimann}},\
  }\bibfield  {title} {\bibinfo {title} {Persistent currents in {B}ose gases
  confined in annular traps},\ }\href
  {https://doi.org/10.1103/PhysRevA.82.043631} {\bibfield  {journal} {\bibinfo
  {journal} {Phys. Rev. A}\ }\textbf {\bibinfo {volume} {82}},\ \bibinfo
  {pages} {043631} (\bibinfo {year} {2010})}\BibitemShut {NoStop}%
\bibitem [{\citenamefont {Kaminishi}\ \emph {et~al.}(2011)\citenamefont
  {Kaminishi}, \citenamefont {Kanamoto}, \citenamefont {Sato},\ and\
  \citenamefont {Deguchi}}]{kaminishi2011}%
  \BibitemOpen
  \bibfield  {author} {\bibinfo {author} {\bibfnamefont {E.}~\bibnamefont
  {Kaminishi}}, \bibinfo {author} {\bibfnamefont {R.}~\bibnamefont {Kanamoto}},
  \bibinfo {author} {\bibfnamefont {J.}~\bibnamefont {Sato}},\ and\ \bibinfo
  {author} {\bibfnamefont {T.}~\bibnamefont {Deguchi}},\ }\bibfield  {title}
  {\bibinfo {title} {Exact yrast spectra of cold atoms on a ring},\ }\href
  {https://doi.org/10.1103/PhysRevA.83.031601} {\bibfield  {journal} {\bibinfo
  {journal} {Phys. Rev. A}\ }\textbf {\bibinfo {volume} {83}},\ \bibinfo
  {pages} {031601(R)} (\bibinfo {year} {2011})}\BibitemShut {NoStop}%
\bibitem [{\citenamefont {Manninen}\ \emph {et~al.}(2012)\citenamefont
  {Manninen}, \citenamefont {Viefers},\ and\ \citenamefont
  {Reimann}}]{manninen2012}%
  \BibitemOpen
  \bibfield  {author} {\bibinfo {author} {\bibfnamefont {M.}~\bibnamefont
  {Manninen}}, \bibinfo {author} {\bibfnamefont {S.}~\bibnamefont {Viefers}},\
  and\ \bibinfo {author} {\bibfnamefont {S.}~\bibnamefont {Reimann}},\
  }\bibfield  {title} {\bibinfo {title} {Quantum rings for beginners {II}:
  Bosons versus fermions},\ }\href
  {https://doi.org/https://doi.org/10.1016/j.physe.2012.09.013} {\bibfield
  {journal} {\bibinfo  {journal} {Physica E: Low-dimensional Systems and
  Nanostructures}\ }\textbf {\bibinfo {volume} {46}},\ \bibinfo {pages} {119 }
  (\bibinfo {year} {2012})}\BibitemShut {NoStop}%
\bibitem [{\citenamefont {Chen}\ \emph {et~al.}(2019)\citenamefont {Chen},
  \citenamefont {Li}, \citenamefont {Proukakis},\ and\ \citenamefont
  {Malomed}}]{chen2019}%
  \BibitemOpen
  \bibfield  {author} {\bibinfo {author} {\bibfnamefont {Z.}~\bibnamefont
  {Chen}}, \bibinfo {author} {\bibfnamefont {Y.}~\bibnamefont {Li}}, \bibinfo
  {author} {\bibfnamefont {N.~P.}\ \bibnamefont {Proukakis}},\ and\ \bibinfo
  {author} {\bibfnamefont {B.~A.}\ \bibnamefont {Malomed}},\ }\bibfield
  {title} {\bibinfo {title} {Immiscible and miscible states in binary
  condensates in the ring geometry},\ }\href
  {https://doi.org/10.1088/1367-2630/ab3207} {\bibfield  {journal} {\bibinfo
  {journal} {New Journal of Physics}\ }\textbf {\bibinfo {volume} {21}},\
  \bibinfo {pages} {073058} (\bibinfo {year} {2019})}\BibitemShut {NoStop}%
\bibitem [{\citenamefont {Bray}\ and\ \citenamefont
  {Simenel}(2021)}]{brayFermionsLongFiniterange2021}%
  \BibitemOpen
  \bibfield  {author} {\bibinfo {author} {\bibfnamefont {A.~W.}\ \bibnamefont
  {Bray}}\ and\ \bibinfo {author} {\bibfnamefont {C.}~\bibnamefont {Simenel}},\
  }\bibfield  {title} {\bibinfo {title} {Fermions with long and finite-range
  interactions on a quantum ring},\ }\href
  {https://doi.org/10.1103/PhysRevC.103.014302} {\bibfield  {journal} {\bibinfo
   {journal} {Phys. Rev. C}\ }\textbf {\bibinfo {volume} {103}},\ \bibinfo
  {pages} {014302} (\bibinfo {year} {2021})}\BibitemShut {NoStop}%
\bibitem [{\citenamefont {Davies}\ \emph {et~al.}(1980)\citenamefont {Davies},
  \citenamefont {Flocard}, \citenamefont {Krieger},\ and\ \citenamefont
  {Weiss}}]{daviesApplicationImaginaryTime1980}%
  \BibitemOpen
  \bibfield  {author} {\bibinfo {author} {\bibfnamefont {K.~T.~R.}\
  \bibnamefont {Davies}}, \bibinfo {author} {\bibfnamefont {H.}~\bibnamefont
  {Flocard}}, \bibinfo {author} {\bibfnamefont {S.}~\bibnamefont {Krieger}},\
  and\ \bibinfo {author} {\bibfnamefont {M.~S.}\ \bibnamefont {Weiss}},\
  }\bibfield  {title} {\bibinfo {title} {Application of the imaginary time step
  method to the solution of the static {{Hartree-Fock}} problem},\ }\href
  {https://doi.org/10.1016/0375-9474(80)90509-6} {\bibfield  {journal}
  {\bibinfo  {journal} {Nucl. Phys. A}\ }\textbf {\bibinfo {volume} {342}},\
  \bibinfo {pages} {111} (\bibinfo {year} {1980})}\BibitemShut {NoStop}%
\bibitem [{\citenamefont {{Mur-Petit}}\ \emph {et~al.}(2002)\citenamefont
  {{Mur-Petit}}, \citenamefont {Polls},\ and\ \citenamefont
  {Mazzanti}}]{mur-loosTwoElectronsHypersphere2009}%
  \BibitemOpen
  \bibfield  {author} {\bibinfo {author} {\bibfnamefont {J.}~\bibnamefont
  {{Mur-Petit}}}, \bibinfo {author} {\bibfnamefont {A.}~\bibnamefont {Polls}},\
  and\ \bibinfo {author} {\bibfnamefont {F.}~\bibnamefont {Mazzanti}},\
  }\bibfield  {title} {\bibinfo {title} {The variational principle and simple
  properties of the ground-state wave function},\ }\href
  {https://doi.org/10.1119/1.1479742} {\bibfield  {journal} {\bibinfo
  {journal} {Am. J. Phys.}\ }\textbf {\bibinfo {volume} {70}},\ \bibinfo
  {pages} {808} (\bibinfo {year} {2002})}\BibitemShut {NoStop}%
\bibitem [{\citenamefont {Cesca}(2022)}]{CescaHonours}%
  \BibitemOpen
  \bibfield  {author} {\bibinfo {author} {\bibfnamefont {J.}~\bibnamefont
  {Cesca}},\ }\emph {\bibinfo {title} {Fermions on a Quantum Ring}},\ \href
  {https://openresearch-repository.anu.edu.au/handle/1885/287311} {\bibinfo
  {type} {Thesis (honours)}},\ \bibinfo  {school} {Australian National
  University} (\bibinfo {year} {2022})\BibitemShut {NoStop}%
\bibitem [{\citenamefont {Needs}\ \emph {et~al.}(2009)\citenamefont {Needs},
  \citenamefont {Towler}, \citenamefont {Drummond},\ and\ \citenamefont
  {L\'opez~R\'{i}os}}]{needs_continuum_2010}%
  \BibitemOpen
  \bibfield  {author} {\bibinfo {author} {\bibfnamefont {R.~J.}\ \bibnamefont
  {Needs}}, \bibinfo {author} {\bibfnamefont {M.~D.}\ \bibnamefont {Towler}},
  \bibinfo {author} {\bibfnamefont {N.~D.}\ \bibnamefont {Drummond}},\ and\
  \bibinfo {author} {\bibfnamefont {P.}~\bibnamefont {L\'opez~R\'{i}os}},\
  }\bibfield  {title} {\bibinfo {title} {Continuum variational and diffusion
  quantum {M}onte {C}arlo calculations},\ }\href
  {https://doi.org/10.1088/0953-8984/22/2/023201} {\bibfield  {journal}
  {\bibinfo  {journal} {J. Phys.: Condens. Matter}\ }\textbf {\bibinfo {volume}
  {22}},\ \bibinfo {pages} {023201} (\bibinfo {year} {2009})}\BibitemShut
  {NoStop}%
\bibitem [{\citenamefont {Metropolis}\ \emph {et~al.}(1953)\citenamefont
  {Metropolis}, \citenamefont {Rosenbluth}, \citenamefont {Rosenbluth},
  \citenamefont {Teller},\ and\ \citenamefont
  {Teller}}]{1953JChemPhysMetropolis}%
  \BibitemOpen
  \bibfield  {author} {\bibinfo {author} {\bibfnamefont {N.}~\bibnamefont
  {Metropolis}}, \bibinfo {author} {\bibfnamefont {A.~W.}\ \bibnamefont
  {Rosenbluth}}, \bibinfo {author} {\bibfnamefont {M.~N.}\ \bibnamefont
  {Rosenbluth}}, \bibinfo {author} {\bibfnamefont {A.~H.}\ \bibnamefont
  {Teller}},\ and\ \bibinfo {author} {\bibfnamefont {E.}~\bibnamefont
  {Teller}},\ }\bibfield  {title} {\bibinfo {title} {Equation of state
  calculations by fast computing machines},\ }\href
  {https://doi.org/10.1063/1.1699114} {\bibfield  {journal} {\bibinfo
  {journal} {J. Chem. Phys.}\ }\textbf {\bibinfo {volume} {21}},\ \bibinfo
  {pages} {1087} (\bibinfo {year} {1953})}\BibitemShut {NoStop}%
\bibitem [{\citenamefont {Flyvbjerg}\ and\ \citenamefont
  {Petersen}(1989)}]{flyvbjergErrorEstimatesAverages1989}%
  \BibitemOpen
  \bibfield  {author} {\bibinfo {author} {\bibfnamefont {H.}~\bibnamefont
  {Flyvbjerg}}\ and\ \bibinfo {author} {\bibfnamefont {H.~G.}\ \bibnamefont
  {Petersen}},\ }\bibfield  {title} {\bibinfo {title} {Error estimates on
  averages of correlated data},\ }\href {https://doi.org/10.1063/1.457480}
  {\bibfield  {journal} {\bibinfo  {journal} {J. Chem. Phys.}\ }\textbf
  {\bibinfo {volume} {91}},\ \bibinfo {pages} {461} (\bibinfo {year}
  {1989})}\BibitemShut {NoStop}%
\bibitem [{\citenamefont {Jastrow}(1955)}]{jastrowManyBodyProblemStrong1955}%
  \BibitemOpen
  \bibfield  {author} {\bibinfo {author} {\bibfnamefont {R.}~\bibnamefont
  {Jastrow}},\ }\bibfield  {title} {\bibinfo {title} {Many-{{Body Problem}}
  with {{Strong Forces}}},\ }\href {https://doi.org/10.1103/PhysRev.98.1479}
  {\bibfield  {journal} {\bibinfo  {journal} {Phys. Rev.}\ }\textbf {\bibinfo
  {volume} {98}},\ \bibinfo {pages} {1479} (\bibinfo {year}
  {1955})}\BibitemShut {NoStop}%
\bibitem [{\citenamefont {Jones}\ \emph {et~al.}(1993)\citenamefont {Jones},
  \citenamefont {Perttunen},\ and\ \citenamefont
  {Stuckman}}]{jonesLipschitzianOptimizationLipschitz1993}%
  \BibitemOpen
  \bibfield  {author} {\bibinfo {author} {\bibfnamefont {D.~R.}\ \bibnamefont
  {Jones}}, \bibinfo {author} {\bibfnamefont {C.~D.}\ \bibnamefont
  {Perttunen}},\ and\ \bibinfo {author} {\bibfnamefont {B.~E.}\ \bibnamefont
  {Stuckman}},\ }\bibfield  {title} {\bibinfo {title} {Lipschitzian
  optimization without the {{Lipschitz}} constant},\ }\href
  {https://doi.org/10.1007/BF00941892} {\bibfield  {journal} {\bibinfo
  {journal} {J. Optim. Theory Appl.}\ }\textbf {\bibinfo {volume} {79}},\
  \bibinfo {pages} {157} (\bibinfo {year} {1993})}\BibitemShut {NoStop}%
\bibitem [{\citenamefont {Johnson}()}]{steven_g_johnson_nlopt_nodate}%
  \BibitemOpen
  \bibfield  {author} {\bibinfo {author} {\bibfnamefont {S.~G.}\ \bibnamefont
  {Johnson}},\ }\href {http://github.com/stevengj/nlopt} {\bibinfo {title} {The
  {NLopt} nonlinear-optimization package}}\BibitemShut {NoStop}%
\end{thebibliography}%

\end{document}